\documentclass[prc,aps,superscriptaddress]{revtex4}  
\usepackage{url,hyperref,lineno,microtype}
\usepackage{bm}
\usepackage{color}
\usepackage{amsfonts}
\usepackage{amssymb}
\usepackage{amsmath}
\usepackage{graphicx}

\definecolor {green}           {rgb}{0.00,0.39,0.00}
\definecolor {blue}            {rgb}{0.00,0.00,1.00}
\definecolor {red}             {rgb}{1.00,0.00,0.00}

\newcommand{\unit}{1\!\!1}
\begin{document}
\newcommand {\beq} {\begin{eqnarray}}
\newcommand {\eeq} {\end{eqnarray}}
\newcommand{\Eq}[1]{Eq.~(\ref{#1})}
\newcommand{\Eqs}[1]{Eqs.~(\ref{#1})}
\newcommand{\Ref}[1]{Ref.~\cite{#1}}
\newcommand{\Refs}[1]{Refs.~\cite{#1}}
\newcommand {\nn} {\nonumber}


\title{The Hyperspherical Harmonics method: a tool for
  testing and improving nuclear interaction models}

\author{L.E.\ Marcucci\,$^{1,2,*}$, J.\ Dohet-Eraly\,$^{3}$,
  L.\ Girlanda\,$^{4,5}$, A.\ Gnech\,$^{2,6}$, A.\ Kievsky\,$^{1,2}$
  and M. Viviani\,$^{1,2}$} 
\affiliation{$^{1}$ Department of Physics ``E. Fermi'', University of Pisa,
  Largo B. Pontecorvo 3, I-56127 Pisa, Italy\\
  $^{2}$ Istituto Nazionale di Fisica Nucleare, Sezione di Pisa,
  Largo B. Pontecorvo 3, I-56127 Pisa, Italy\\
  $^{3}$ Physique Quantique, and Physique Nucl\'eaire Th\'eorique et
  Physique Math\'ematique, C.P. 229, Universit\'e libre de Bruxelles (ULB),
  B-1050 Brussels, Belgium\\
  $^{4}$ Department of Mathematics and Physics Matematica ``E. De Giorgi'',
  University of Salento, Ex Collegio Fiorini - Via per Arnesano,
  I-73100 Lecce, Italy\\
  $^{5}$ Istituto Nazionale di Fisica Nucleare, Sezione di Lecce,
  Via per Arnesano, I-73100 Lecce, Italy\\
  $^{6}$ Gran Sasso Science Institute, Viale Francesco Crispi 7,
  I-67100 L'Aquila, Italy} 

\begin{abstract}

The Hyperspherical Harmonics (HH) method is one of the
most accurate techniques to solve the quantum mechanical problem
for nuclear systems with $A\le 4$. In particular,
by applying the Rayleigh-Ritz or Kohn variational principle,
both bound and scattering states can be addressed, using either
local or non-local interactions. Thanks to this
versatility, the method can be used to test the two- and
three-nucleon components of the nuclear interaction. 

In the present review we introduce the formalism of the HH method,
both for bound and
scattering states. In particular, we describe the
implementation of the method to study the $A=3$ and $4$
scattering problem.
Second, we present a selected choice of results of the last decade,
most representative of the latest achievements.
Finally, we conclude with a discussion of
what we believe will be the most significant
developments within the HH method for the next
five-to-ten years.

\end{abstract}

\maketitle
\section{Introduction}
\label{sec:intro}

The ``standard'' picture of a nucleus sees it as a system of $A$ nucleons,
protons or neutrons, interacting among themselves and eventually with
external electro-weak probes. The interaction between nucleons, i.e.\
the nuclear interaction, is the subject of the Research Topic of
which this contribution is part. Using the nucleon as the relevant
degree of freedom, the system is described by the nuclear
Hamiltonian, written as
\begin{equation}
  H=\sum_i \frac{p_i^2}{2m_i}+\sum_{i<j} V_{ij}+\sum_{i<j<k} V_{ijk} +\cdots \ ,
  \label{eq:nuclH}
\end{equation}
where the first term is the (non-relativistic) kinetic energy, $m_i$
being the $i$th nucleon mass,
$V_{ij}$ and $V_{ijk}$ are, respectively, the two- and three-nucleon
interactions, i.e.\ the interaction between each $ij$ pair or $ijk$ triple.
It has been shown in several studies (for recent ones
see Refs.~\cite{Piarulli:2018,Epelbaum:2019}) that even the
nuclear systems with medium-large values of $A$ can be
at least qualitatively described including $V_{ij}$ and $V_{ijk}$ only:
essentially, it seems to be no room for four- or more-nucleon
interactions (the dots of Eq.~(\ref{eq:nuclH})).
Therefore, we will neglect from now on
the contributions beyond three-nucleon interaction.

There exists a large variety of realistic models
for $V_{ij}$ and $V_{ijk}$, and the most important ones
are presented in this Research Topic. These models are
very different among themselves, as they can be local,
or minimally non-local and expressed in coordinate space,
or non-local and expressed in momentum space.
Some older models were derived phenomenologically or in a
meson-theoretical approach, as the Argonne $v_{18}$~\cite{Wiringa:1995}
or the CDBonn~\cite{Machleidt:2001} potentials, but, since
the seminal work of Weinberg~\cite{Weinberg:1990}, the
preferred framework to derive the nuclear interaction
is chiral effective field theory. Since the presentation
of the different models is assigned to this Research Topic,
here we only mention that all the models have a common characteristics: the
$V_{ij}$ and $V_{ijk}$ interactions have an intricate operatorial structure.
As a consequence, the solution of the Schr\"odinger equation for $A>2$
becomes a challenging problem. Those methods which are
able to solve the $A$-body quantum mechanical problem without approximations,
or with approximations which can be maintained under control, are
the so-called {\it ab-initio} methods~\footnote{The expression ``{\it ab-initio}
  method'' has been quite widely used in the literature, with a somewhat
  less strict definition, than the one used here. Here we follow
  the definition of Ref.~\cite{Leidemann:2013}.}. They
are clearly essential in order to test a given model
for the nuclear interaction against experiment. It is
fundamental for these methods to be sufficiently accurate
to capture the small details introduced by the complexity of the interaction. 
As an example, we can quote the case of the triton binding energy. It is
nowadays well known that the triton binding energy calculated
just retaining $V_{ij}$ in Eq.~(\ref{eq:nuclH})
is underestimated by $0.5-1$ MeV,
depending of the considered model.
It is straightforward that the required accuracy
of the {\it ab-initio} method to catch this disagreement
must be better that 500 keV. Nowadays, the methods for the
$A=3$ bound systems have reached a much higher accuracy, of the
order of 1 keV, or even better. And therefore, the presence
of $V_{ijk}$ is not anymore under discussion.

There are several {\it ab-initio} methods which can
solve the $A$-body quantum mechanical problem in different
regions of the nuclear chart. A recent review is given in
Ref.~\cite{Leidemann:2013}. Here we limit ourselves to mention
the methods based on Monte Carlo techniques, as the
variational Monte Carlo (VMC) or the Green's function Monte Carlo (GFMC)
methods (see Ref.~\cite{Carlson:2015}, and references therein).
There are then the methods linked to the shell model,
as the no-core shell model (NCSM) or the realistic shell model (RSM),
see Refs.~\cite{Barrett:2013} and~\cite{Fukui:2018,Coraggio:2012},
respectively. All these methods are quite
powerful to study medium-mass nuclear bound states, but less accurate, apart
from the GFMC and NCSM, for very light nuclei, as those with $A=3,4$.
Furthermore, their extension to the scattering systems is not so
trivial, and, in some cases, still not at reach.

Restricting ourselves to the $A=3,4$ nuclear systems, both 
bound and scattering states, we have at hand very
few accurate {\it ab-initio} methods, i.e.\ the Faddeev (Faddeev-Yakubovsky
for $A=4$) equations (FE) technique, solved
in coordinate- or in momentum-space, the method
based on the Alt-Grassberger-Sandhas (AGS) equations solved in momentum
space, and the Hyperspherical Harmonics (HH) method presented here.
We refer the reader to Refs.~\cite{Lazauskas:2004,Lazauskas:2019}
for the FE method in
coordinate space, to Refs.~\cite{Glockle:1996,Nogga:1997}
for the FE method in momentum space, to
Refs.~\cite{Fonseca:2017,Deltuva:2019} for recent reviews on
the AGS method. Clearly,
each method has advantages and drawbacks. For instance,
the FE method in momentum space can be applied to $A=3,4$ bound
and scattering states in a wide energy range. However, the inclusion
of the Coulomb interaction for charged particle scattering states
is quite problematic. The FE method in coordinate space
can handle the Coulomb interaction, but it has not yet been applied
to scattering problems at very low-energy, and it has been applied
only recently to study systems with larger $A$
values~\cite{Lazauskas:2018}. It is though a method with in principle
great possibilities of extension~\cite{Lazauskas:2019}. The AGS method,
although working in momentum space, can handle the Coulomb interaction,
and can be applied to a large variety of $A=3,4$ 
scattering states, in a wide energy range. However, the very low
energy range, that of interest for nuclear astrophysics, i.e.\
below $\sim 100$ keV, is still not accessible with the AGS method.
The method has also not been applied for $A>4$ yet.

The HH method has a long history, nicely summarized in the introduction
of Ref.~\cite{Kievsky:2008}. We will concentrate here on the
latest developments, essentially those obtained
since the year of 2008, year of publication
of Ref.~\cite{Kievsky:2008}. However, to fully appreciate the great
developments of this last decade, it is necessary to briefly
outline the state-of-the-art of the HH method at that time. The HH method
in 2008 existed in two flavours: the correlated HH method, including a
pair-correlation function (pair-correlated HH method -- PHH)
or with a Jastrow type factor (correlated HH method -- CHH),
and the ``pure'' HH method. The correlation factor was introduced
to describe correlations induced by the interaction at short ranges.
In fact, when the interaction has a strong repulsion at short range,
the wave function goes to zero when two particles are close to each other.
The correlation factor describes this particular configuration
and goes to unity for large pair relative distances. 
Therefore the HH expansion has to take care of reconstructing 
the wave function outside the range in which the interaction shows the
strong repulsion, making the convergence of the expansion much faster. 
The drawbacks
of the PHH (or CHH) method are (i) the necessity of performing numerical
integrations, which would be instead analytical without correlation factors,
that could reduce the accuracy of the method in the $A=4$ case; (ii) the not
simple extension of the PHH method to work in momentum space. Therefore,
it is difficult to apply the PHH method with the non-local potentials
mentioned above. This has motivated us, together with the continous
increasing of computing powers, to return to the ``pure'' HH method.
Up to the year 2008, this had been developed and applied
to study with great accuracy the $A=3,4$ bound states,
with both local potentials, expressed in coordinate space,
or non-local ones, given in momentum-space. In fact, while the
local interactions had been at reach for the HH method from the
very beginning~\cite{Kievsky:1997a}, the
non-local ones were a recent achievement at that time~\cite{Viviani:2006}.
In 2008, the zero-energy $A=3,4$ scattering states were also
calculated with both local and non-local interactions~\cite{Kievsky:2008}.
The higher energy scattering states, still below the breakup
threshold of the target nucleus, had been studied for both $A=3$ and
$4$ systems only with local interactions, in a variety of contributions
extensively mentioned in Ref.~\cite{Kievsky:2008}. What was still missing
in 2008 was the study of the $A=3,4$ scattering states, still below
the target breakup threshold, with non-local potentials. This has
been obtained in Refs.~\cite{Marcucci:2009,Viviani:2008,Viviani:2010,
  Viviani:2011,Viviani:2013} for both $A=3$ and $4$, and it is
in fact one of the main achievements of the HH method of this last decade.
The HH method, in its PHH version,
has been applied to describe elastic scattering observables in $A=3$
above the deuteron breakup threshold~\cite{Kievsky:2001} and
in wide energy region
including the full electromagnetic interaction~\cite{Kievsky:2004}. 
Preliminary studies of the method to treat the breakup channels, as for
instance to the
process $n+d\rightarrow n+n+p$, can be found in 
Refs.~\cite{Kievsky:1999,Viviani:2001b,Garrido:2014}.
The application of the method using the Hamiltonian defined in
Eq.~(\ref{eq:nuclH}) is in progress.

A further development of the method is toward larger values of $A$.
This has been performed within the so-called non-symmetrized HH
method~\cite{Gattobigio:2011} with central potentials.
The implementation of the HH method to $A\le 6$ using realistic interactions
has been performed within the so-called effective interaction HH method
by the Trento group (see Ref.~\cite{Leidemann:2013} and references
therein), with a procedure, the Suzuki-Lee
approach~\cite{Okubo:1954,Suzuki:1982,Suzuki:1983},
which significantly reduces the number of the
basis functions needed in the expansion. The first steps to
use the HH method without the Suzuki-Lee approach have been shown in
Ref.~\cite{Dohet-Eraly:2019}, and intense research activity is
currently underway. The formalism which will be presented here
is in fact quite general, and can be applied also to the $A=5,6$ nuclear
systems.

Before concluding this section with the outline of this contribution,
we would like to make few remarks: (i) the HH method is extremely
powerful, and its application to systems up to $A\sim 7,8$
is limited essentially by computing power. (ii) The accuracy of the HH method has been tested
in a number of benchmark calculations. In particular we quote
the benchmark on the $A=3$~\cite{Nogga:2003} and
$A=4$~\cite{Kamada:2001} bound states, on the $nd$ and $pd$ scattering
phase shifts~\cite{Kievsky:1998,Deltuva:2005}, and, in the last decade,
also on the
$A=4$ scattering states~\cite{Viviani:2011,Viviani:2016}.
(iii) Compared with the other {\it ab-initio} methods,
the HH technique seems to be one of the best choices
to study low-energy scattering states, in order to obtain
accurate predictions for nuclear reactions of astrophysical
interest~\cite{Marcucci:2006,Marcucci:2016}.

The present review is organized as follows: in Sec.~\ref{sec:formalism}
we discuss the HH formalism, both for bound and scattering states.
We will try to keep a somewhat ``pedagogical'' level, in order to
allow the interested reader to perform his/her own
algebraic steps and eventually reproduce the already existing
results. In Sec.~\ref{sec:res} we discuss the most important
results obtained within the HH method since the year 2008.
In particular, we will show that the HH method has reached
such a degree of accuracy for both bound and scattering states,
that it has been used in order to construct an accurate model
for the three-nucleon interaction, with a procedure similar,
in principle, to the one used to derive the nowadays very accurate
two-nucleon interaction models. Finally, in Sec.~\ref{sec:concl-out}
we will give some concluding remarks and an outlook.

\section{The HH formalism}
\label{sec:formalism}
We review in this section the HH formalism, focusing in particular
on the three- and four-body systems, both bound and scattering states.
The approach described can be used in conjunction with both
local and non-local two-nucleon interactions. At present, the method
works with only local three-nucleon interactions, but its extension to
the non-local case does not lend to conceptual difficulties.

\subsection{Hyperspherical harmonic functions}
\label{subsec:HHfunctions}

Let us consider a system of $A$ particles with masses $m_1$, $\ldots$, $m_A$
and spatial coordinates $\bm{r}_1$, $\ldots$, $\bm{r}_A$, respectively. For
separating the internal and center-of-mass (c.m.) motion, it is convenient to
introduce another set of coordinates made of $N=A-1$ internal Jacobi
coordinates $\bm{x}_1$, $\ldots$, $\bm{x}_N$ and the c.m.\ coordinate $\bm{X}$
defined by  
\beq
\bm{X}=\dfrac{1}{M} \sum^A_{i=1} m_i\bm{r}_i \ ,
\label{eq:defX}
\eeq
where $M=\sum^A_{i=1}m_i$ is the total mass of the system.
There are several definitions of the Jacobi coordinates, but a convenient
one which will be used through this work is
the following
\beq\label{eq:defJac}
\bm{x}_{N-j+1}=\sqrt{\dfrac{2 m_{j+1} M_j}{M_{j+1} m}}
\left(\bm{r}_{j+1}-\dfrac{1}{M_j} \sum^j_{i=1} m_i \bm{r}_i\right) \ ,
\eeq
where $m$ is a reference mass, $M_j=\sum^j_{i=1} m_i$, and $j=1,\ldots,N$.
In the case where all the particles have the same mass $m$, \Eq{eq:defJac}
reduces to 
\beq\label{eq:defJac_equalMass}
\bm{x}_{N-j+1}=\sqrt{\dfrac{2 j}{j+1}} \left(\bm{r}_{j+1}-\dfrac{1}{j}
\sum^j_{i=1} \bm{r}_i \right) \ .
\eeq
From a given choice of the Jacobi vectors,
the hyperspherical coordinates $(\rho,\Omega_N)$ can be introduced.
The hyperradius $\rho$ is defined by
\beq\label{eq:defrho}
\rho=\sqrt{\sum^N_{i=1} x^2_i}
=\sqrt{\frac{2}{A}\sum^A_{j>i=1} (\bm{r}_i-\bm{r}_j)^2}=
\sqrt{2 \sum^A_{i=1}(\bm{r}_i-\bm{X})^2} \ ,
\eeq
where $x_i$ is the modulus of the Jacobi vector $\bm{x}_i$.
The hyperradius $\rho$ 
is symmetric with respect to particle exchanges
and does not depend on the particular choice of Jacobi coordinates.
The set $\Omega_N$ of hyperangular coordinates,
\beq\label{eq:defOmega}
\Omega_N=\{\hat{\bm x}_1,\ldots,\hat{\bm x}_N,\varphi_2,\ldots,\varphi_N\},
\eeq
is made of the angular parts $\hat{\bm x}_i=(\theta_i,\phi_i)$ of the
spherical components of the Jacobi vectors ${\bm{x}}_i$,
and of the hyperangles $\varphi_i$,
defined by
\beq\label{eq:defHyperAngle}
\cos\varphi_i=\dfrac{x_i}{\sqrt{x^2_1+\ldots+x^2_i}} \ ,
\eeq
where $0\leq \varphi_i\leq \pi/2$
and $i=2,\ldots,N$. 

The advantage of using the hyperspherical coordinates can be appreciated
noting that the internal kinetic energy operator of the $A$-body system
can be decomposed as
\beq\label{eq:laplac}
T=-\dfrac{\hbar^2}{m}\sum_{i=1}^N\Delta_{{\bf x}_i}=
  -\dfrac{\hbar^2}{m}\left(\frac{\partial^2}{\partial\rho^2}+
  \frac{3N-1}{\rho}\frac{\partial}{\partial\rho}-
  \frac{\Lambda_N^2(\Omega_N)}{\rho^2}\right),
\eeq
where the operator $\Lambda_N^2(\Omega_N)$ is the so-called
grand-angular momentum operator. Its explicit expression can be found,
for instance,
in \Refs{Ripelle:1983,Kievsky:2008}, but it is not
essential for our purposes.
More important are the eigenfunctions of the grand-angular
momentum $\Lambda_N^2(\Omega_N)$, the so-called hyperspherical harmonics
(HH). They can be defined as
\beq\label{eq:defHH}
\mathcal{Y}^{KLM_L}_{[K]}(\Omega_N)=[[\ldots[Y_{l_1}(\hat{\bm x}_1)
      Y_{l_2}(\hat{\bm x}_2)]_{L_2}
    \ldots Y_{l_{N-1}}(\hat{\bm x}_{N-1})]_{L_{N-1}} Y_{l_N}(\hat{\bm x}_N)]_{LM_L}
\prod^N_{j=2}
{^{(j)}\mathcal{P}}^{K_{j-1},l_j}_{n_j}(\varphi_j)\ .
\end{eqnarray}
Here $Y_{l_i}(\hat{\bm x}_i)$ is a spherical harmonic function
for $i=1,\ldots, N$, $L$ is the total orbital angular momentum, $M_L$
its projection on the $z$ axis,
and 
\beq\label{eq:Kj}
K_j&=&\sum^j_{i=1} (l_i+2 n_i)
\eeq
with $n_1= 0$, $j=1,\ldots,N$, and $K_N\equiv K$ is the so-called
grand-angular momentum.
The notation $[K]$ stands for the collection of all the quantum
numbers $[l_1,\ldots,l_N,L_2,\ldots,L_{N-1},n_2,\ldots,n_N]$.
The functions ${^{(j)}\mathcal{P}}^{K_{j-1},l_j}_{n_j}(\varphi_j)$
in \Eq{eq:defHH} are defined by
\beq
    {^{(j)}\mathcal{P}}^{K_{j-1},l_j}_{n_j}=\mathcal{N}^{l_j,\nu_j}_{n_j}
    (\cos\varphi_j)^{l_j} (\sin\varphi_j)^{K_{j-1}}
    P^{\nu_{j-1},l_j+1/2}_{n_j}(\cos 2\varphi_j),
\eeq
where  $P^{\nu_{j-1},l_j+1/2}_{n_j}(\cos 2\varphi_j)$
are Jacobi polynomials~\cite{Abramowitz}, with
\beq\label{eq:nuj}
\nu_j=K_j+\dfrac{3}{2}j-1 \ ,
\eeq
and the normalization factors $\mathcal{N}^{l,\nu}_{n}$ are given by
\beq\label{eq:N}
\mathcal{N}^{l,\nu}_{n}=\sqrt{\frac{2 \nu \Gamma(\nu-n)\Gamma(n+1)}{\Gamma(\nu-n-l-1/2)\Gamma(n+l+3/2)}},
\eeq
with $\Gamma$ indicating the standard Gamma function~\cite{Abramowitz}.
With the definition of \Eq{eq:defHH}, the HH functions are 
eigenvectors of the grand-angular momentum operator $\Lambda^2_N (\Omega_N)$,
the square of the total orbital angular momentum $\bm{L}$,
its $z$ component ${\bm{L}}_z$, and the parity operator $\Pi$. Therefore
we have
\beq
\Lambda^2_N(\Omega_N)\mathcal{Y}^{KLM_L}_{[K]}(\Omega_N)&=&
K(K+3N-2)\mathcal{Y}^{KLM_L}_{[K]}(\Omega_N) \ , \label{eq:propLambda}\\
L^2\mathcal{Y}^{KLM_L}_{[K]}(\Omega_N)&=&\hbar^2 L(L+1)
\mathcal{Y}^{KLM_L}_{[K]}(\Omega_N) \ , \label{eq:propL2}\\
L_z \mathcal{Y}^{KLM_L}_{[K]}(\Omega_N)&=&\hbar
M_L \mathcal{Y}^{KLM_L}_{[K]}(\Omega_N) \ , \label{eq:propLz}\\
\Pi\mathcal{Y}^{KLM_L}_{[K]}(\Omega_N)
&=&(-1)^{K}\mathcal{Y}^{KLM_L}_{[K]}(\Omega_N) \ . \label{eq:propPi}
\eeq
We remark here two useful properties of the HH functions. First
of all, the HH functions are orthonormal with respect to the volume
element $d\Omega_N$, i.e.
\beq\label{eq:HHortho}
\int d\Omega_N\,
[\mathcal{Y}^{K'L'M_L'}_{[K']}(\Omega_N)]^*\,
\mathcal{Y}^{KLM_L}_{[K]}(\Omega_N)=\delta_{[K] [K']}
\delta_{K K'}\delta_{L L'}\delta_{M_L M_L'} \ ,
\eeq
with
\beq\label{eq:d3xi}
d{\bm x}_1\cdots d{\bm x}_N=\rho^{D-1}d\rho \, d\Omega_N
\eeq
and
\beq\label{eq:dOmega}
d\Omega_N=\sin\theta_1d\theta_1d\phi_1
  \prod_{j=2}^N \sin\theta_jd\theta_jd\phi_j\,(\cos\varphi_j)^2\,
  (\sin\varphi_j)^{3j-4}d\varphi_j \ .
\eeq
Therefore, the number of HH functions for a given $K$ increases fast
with $K$, but is always finite. In fact, according with Eq.~(\ref{eq:Kj}),
  $K=\sum_{i=1,N}(l_i+2n_i)$.
Furthermore, independently of 
the specific choice of Jacobi coordinates used to define the
hyperspherical ones or of the order of the coupling of the spherical harmonics
in \Eq{eq:defHH}, the HH functions constitute a complete basis. 

Secondly, in order to evaluate matrix elements of a given
many-body operator between HH functions, it is often useful to determine
the effect of a particles permutation on an HH function. Since the
grand-angular and the total orbital angular momenta are fully symmetric,
and since the HH functions constitute a complete basis, the permuted HH
functions $\mathcal{Y}^{KLM_L}_{[K]}(\Omega_N^p)$ can be written as linear
combinations of unpermuted HH functions $\mathcal{Y}^{KLM_L}_{[K']}(\Omega_N)$
with same $K$, $L$, and $M_L$ values. Therefore, we can write
\beq\label{eq:defTC}
\mathcal{Y}^{KLM_L}_{[K]}(\Omega_N^p)=\sum_{[K']} a^{KL,p}_{[K];[K']}
\mathcal{Y}^{KLM_L}_{[K']}(\Omega_N) \ .
\eeq
%
The transformation coefficients $a^{KL,p}_{[K];[K']}$ do not
depend on the quantum number $M_L$. For $A=3$,  they are called the
Raynal-Revai coefficients~\cite{Raynal:1970}. To be remarked that
$[K']\equiv[l'_1,\ldots,l'_N,L'_2,\ldots,L'_{N-1},n'_2,\ldots,n'_N]$, but
such that $K'=K$. Note that also $L'=L$, i.e.\ $L$ is conserved.
For $A>3$, see Refs.~\cite{Viviani:1998,Dohet-Eraly:2019}.

Let us consider more specifically a system of $A$ nucleons described within
the isospin formalism. The $A$-nucleon wave function contains spatial, spin,
and isospin parts. We can define the spin functions $\chi^{SM_S}_{[S]}$ with
total spin $S$ and total spin projection $M_S$ and the isospin functions
$\xi^{TM_T}_{[T]}$ with total isospin $T$ and total isospin projection $M_T$ by
coupling the individual spin functions $\chi_{1/2,\pm 1/2}$ or isospin
functions $\xi_{1/2,\pm 1/2}$, respectively, of each nucleon, as
\beq
\chi^{SM_S}_{[S]}&=&[[\ldots[\chi_{1/2}(1)\chi_{1/2}(2)]_{S_2}\ldots
    \chi_{1/2}(N-1)]_{S_{N-1}}\chi_{1/2}(N)]_{SM_S} \ , \label{eq:chiS} \\
\xi^{TM_T}_{[T]} &=& [[\ldots[\xi_{1/2}(1)\xi_{1/2}(2)]_{T_2}\ldots
    \xi_{1/2}(N-1)]_{T_{N-1}}\xi_{1/2}(N)]_{TM_T} \ . \label{eq:xiT}
\eeq
So now $[S]$ stands for $[S_2,\ldots,S_{N-1}]$ and $[T]$ for
$[T_2,\ldots,T_{N-1}]$. 

Including the spin and isospin functions, the HH basis functions read
\beq\label{def:HHST}
\mathbb{Y}^{KLSJJ_z T M_T}_{[KST]}(\Omega_N)=[\mathcal{Y}^{KL}_{[K]}(\Omega_N)\chi^{S}_{[S]}]_{J J_z}\xi^{TM_T}_{[T]} \ ,
\eeq
where $J$ is the total angular momentum, $J_z$ its projection, and
$[KST]$ stands for $[K][S][T]$. To be noticed that also the spin-isospin
part of $\mathbb{Y}^{KLSJJ_z T M_T}_{[KST]}(\Omega_N)$ constructed with a given
ordering of the particles, can be rewritten in terms of a different
permutation, using the Wigner 6j coefficients~\cite{Edmonds}.

We conclude by noting that the HH functions can also be built in momentum
space instead of configuration space. They can be obtained by replacing the
hyperspherical coordinates $(\rho,\Omega_N)$ associated with the Jacobi
coordinates $\{\bm{x}_i\}_{i=1,\ldots,N}$ by the hyperspherical coordinates
$(Q,\Omega_N^{(q)})$ associated with the $N$ Jacobi conjugate momenta
$\{\bm{q}_i\}_{i=1,\ldots,N}$. The rest of the formalism remains unchanged.
For more details, see Refs.~\cite{Rosati:2002,Viviani:2006,Kievsky:2008}.

\subsection{The HH method for $A=3$ and $4$}
\label{subsec:a34}
We discuss in some detail the method for systems with $A=3,4$ nucleons
within the isospin formalism for both bound and scattering
states in Sec.~\ref{subsubsec:a34_bound} and~\ref{subsubsec:a34_scatt},
respectively. The extension to $A>4$ is straigthforward, but leads to more
lengthy expressions.

\subsubsection{The $A=3$ and $4$ bound states}
\label{subsubsec:a34_bound}

The wave function of an $A$-body bound state, with $A=3,4$, having
total angular momentum $J,J_z$ and parity $\pi$, and third component of
the total isospin $M_T$, can be decomposed as a sum of Faddeev-like
amplitudes as:
\begin{equation}
  \Psi_A=\sum_{p=1}^{N_p} \psi({\bm x}_1^{(p)}, \cdots, {\bm x}_N^{(p)})
  \label{eq:PsiA} \ .
\end{equation}
Here the sum on $p$ runs up to $N_p=3$ or 12 even permutations of the
$A$ particles, with $A=3$ or $4$, and the coordinates
${\bm x}_1^{(p)}, \cdots, {\bm x}_N^{(p)}$ are the Jacobi coordinates
as defined in \Eq{eq:defJac}. To be noticed that, increasing the number
of particles, different arrangements of them in sub-clusters allow for different
definitions of the Jacobi coordinates. For example, in $A=4$ two different
sets exist corresponding to have a 3+1 or a 2+2 asymptotic configuration. 
However in the sub-space defined by the grand angular momentum $K$,
HH functions defined in different sets of Jacobi coordinates
result to be linearly 
dependent. In the following we always refer to the set defined in
\Eq{eq:defJac}.

The coordinate-space hyperspherical coordinates are given
in \Eqs{eq:defrho}--~\eqref{eq:defHyperAngle}, and the hyperangular
variables are $\varphi_2$ for $A=3$ and $\varphi_2, \varphi_3$ for $A=4$.

We rewrite here the HH basis of \Eq{def:HHST} for the $A=3$ and $4$ case.
Historically, the angular, spin and isospin quantum numbers have been
collected in the so-called channels $\alpha$,
defined explicitly by
\beq
  [\alpha]&=& [l_{1\alpha}, l_{2\alpha}, L_\alpha, S_{a\alpha}, S_\alpha,
  T_{a\alpha}, T_\alpha]\ ; \,\,\,\,\,\,\,\,\, A=3 \label{eq:alpha3} \\ \relax
  [\alpha]&=& [l_{1\alpha}, l_{2\alpha}, l_{3\alpha}, L_{2\alpha}, L_\alpha,
  S_{a\alpha}, S_{b\alpha}, S_\alpha,
  T_{a\alpha}, T_{b\alpha}, T_\alpha]\ ; \,\,\,\,\,\,\, A=4 \label{eq:alpha4}
\eeq
so that we can write
\beq
\mathbb{Y}^{K}_{[\alpha] n_2}(\Omega_N)&=&\Bigg[
    [Y_{l_{1\alpha}}({\hat{\bm x}}_1)\,Y_{l_{2\alpha}}({\hat{\bm x}}_2)]_{L_\alpha}
     \Big[
      [\chi_{1/2}(1)\,\chi_{1/2}(2)]_{S_{a\alpha}}\,\chi_{1/2}(3)\Big]_{S_\alpha}
    \Bigg]_{J J_z} \, \nonumber \\
  &&  \Big[
    [\xi_{1/2}(1)\,\xi_{1/2}(2)]_{T_{a\alpha}}\,\xi_{1/2}(3)\Big]_{T_\alpha M_T}
  \,\,  ^{(2)}{\mathcal P}_{n_2}^{l_{1\alpha}, l_{2\alpha}}(\varphi_2)\ , 
    \label{eq:HH3}
\end{eqnarray}
for $A=3$, and
\beq
\mathbb{Y}^{K}_{[\alpha] n_2 n_3}(\Omega_N)&=&\Bigg[
    \Big[
      [Y_{l_{1\alpha}}({\hat{\bm x}}_1)\,Y_{l_{2\alpha}}({\hat{\bm x}}_2)]_{L_{2\alpha}}
      Y_{l_{3\alpha}}({\hat{\bm x}}_3)\Big]_{L_\alpha} \nonumber \\
    && \Big[\Big[
        [\chi_{1/2}(1)\,\chi_{1/2}(2)]_{S_{a\alpha}}\,\chi_{1/2}(3)\Big]_{S_{b\alpha}}\,
      \chi_{1/2}(4) \Big]_{S_\alpha}\Bigg]_{J J_z}
  \,\nonumber \\
&&  \Big[\Big[
      [\xi_{1/2}(1)\,\xi_{1/2}(2)]_{T_{a\alpha}}\,\xi_{1/2}(3)\Big]_{T_{b\alpha}}\,
    \xi_{1/2}(4) \Big]_{T_\alpha M_T}
  \nonumber \\
&&  ^{(2)}{\mathcal P}_{n_2}^{l_{1\alpha}, l_{2\alpha}}(\varphi_2)
  ^{(3)}{\mathcal P}_{n_3}^{2n_2+l_{1\alpha}+l_{2\alpha}, l_{3\alpha}}(\varphi_3)
  \ ,  
    \label{eq:HH4}
\eeq
for $A=4$. 
To be noticed that, in order to ensure the antisymmetry of the
wave function, the Faddeev-like amplitudes have to change sign under
exchange of particle $1$ and $2$. Therefore, the sum
$l_{2\alpha}+S_{a\alpha}+T_{a\alpha}$ for $A=3$ and
$l_{3\alpha}+S_{a\alpha}+T_{a\alpha}$ for $A=4$ must be odd.
Furthermore, $l_{1\alpha}+l_{2\alpha}$ for $A=3$ and
$l_{1\alpha}+l_{2\alpha}+l_{3\alpha}$ for $A=4$ must be an even or odd number
in correspondence to a positive or negative parity state.
Even with these restrictions, there is an infinite number of channels.
However, the contributions of the channels with higher and higher
values for $l_{1\alpha}+l_{2\alpha}$ for $A=3$ and
$l_{1\alpha}+l_{2\alpha}+l_{3\alpha}$ for $A=4$ should become less and less
important, due to the centrifugal barrier. Therefore,
it is found that the number of channels with a significant
contribution is relatively small. 
The most important ones for $A=3$
and for $A=4$ are listed in
Tables~1 and~2 of Ref.~\cite{Kievsky:2008}, respectively.

By using \Eqs{eq:HH3} and~\eqref{eq:HH4}, 
the $A$-body wave function
$\Psi_A$ of Eq.~\eqref{eq:PsiA} can be written
in coordinate-space as
\begin{equation}
  \Psi_A=\sum_{\alpha,n_2}\,u_{\alpha n_2}(\rho)\sum_p
  \mathbb{Y}^K_{[\alpha] n_2}(\Omega_2^{(p)}) \ , \label{eq:PsiA-HH3}
\end{equation}
for $A=3$, and
\begin{equation}
  \Psi_A=\sum_{\alpha,n_2,n_3}\,u_{\alpha n_2 n_3}(\rho)\sum_p
  \mathbb{Y}^K_{[\alpha] n_2 n_3}(\Omega_3^{(p)}) \ , \label{eq:PsiA-HH4}
\end{equation}
for $A=4$. The sum over $n_2$ in Eq.~(\ref{eq:PsiA-HH3}) and $n_2,n_3$
in Eq.~(\ref{eq:PsiA-HH4}) is restricted to independent states, see below.
The hyperradial functions $u_{\alpha n_2}(\rho)$ ($u_{\alpha n_2 n_3}(\rho)$ for
$A=4$) are themselves expanded in terms of known functions.
It is common to use Laguerre polynomials, as they have been found
to give a stable and nice convergence of this expansion.
Therefore,
\begin{equation}
  u_{\alpha n_2/\alpha n_2 n_3}(\rho)=\sum_m c_{\alpha n_2/\alpha n_2 n_3;m} f_m(\rho) \ ,
  \label{eq:urho}
\end{equation}
where the functions $f_m(\rho)$ are written as
\begin{equation}
  f_m(\rho)=\gamma^{D/2}\sqrt{\frac{m!}{(m+D-1)!}}\,\,L_m^{(D-1)}(\gamma\rho)\,\,
  {\rm e}^{-\gamma\rho/2} \ .
  \label{eq:Laguerre}
\end{equation}
Here $D\equiv 3N-1$,
$L_m^{(D-1)}(\gamma\rho)$ is a Laguerre polynomial~\cite{Abramowitz},
and $\gamma$ is a non-linear parameter, to be variationally optimized.
The exponential factor ${\rm e}^{-\gamma\rho/2}$ ensures that
$f_m(\rho)\rightarrow 0$ for $\rho\rightarrow\infty$. The optimal value
of $\gamma$ depends on the potential model, and it is typically
in the interval 2.5--4.5 fm$^{-1}$ for local and
4--8 fm$^{-1}$ for non-local potentials.

When working in momentum space, the $A$-body wave function
$\Psi_A$ is written as in \Eqs{eq:PsiA-HH3} and~\eqref{eq:PsiA-HH4},
with $u_{\alpha n_2}(\rho)$ and $u_{\alpha n_2 n_3}(\rho)$ replaced
with $w_{\alpha n_2}(Q)$ and $w_{\alpha n_2 n_3}(Q)$, i.e.\ functions
of the hypermomentum $Q$, while the HH functions are expressed
in terms of conjugate Jacobi momenta.
The $w$-functions are related to the
$u$-functions as
\begin{equation}
  w_{\alpha n_2/\alpha n_2 n_3}(Q)=(-i)^K\int_0^\infty d\rho
  \frac{\rho^{D-1}}{(Q\rho)^{D/2-1}}\,J_{K+\frac{D}{2}-1}(Q\rho)\,
  u_{\alpha n_2/\alpha n_2 n_3}(\rho)
  \label{eq:wu-rel} \ ,
\end{equation}
where $J_{K+\frac{D}{2}-1}(Q\rho)$ are Bessel functions of the first
kind~\cite{Kievsky:2008}, and $K$ is again the grand-angular
momentum.

At the end, the $A$-body wave function
of \Eqs{eq:PsiA-HH3}--\eqref{eq:wu-rel}
can be cast in the form
\begin{equation}
  \Psi_A=\sum_{K,m}c_{K;m} |K,m\rangle \ ,
  \label{eq:PsiA_final}
\end{equation}
where
\begin{equation}
  |K,m\rangle\equiv f_m(\rho) \sum_p
  \mathbb{Y}^K_{[\alpha] n_2/[\alpha] n_2 n_3}(\Omega_N^{(p)})
  \ , \label{eq:Gm-r}
\end{equation}
in coordinate-space (a similar expression holds in momentum-space).
The decomposition proposed in \Eq{eq:PsiA}
ensures the complete antisymmetrization of the
state through the sum on the permutations as indicated in \Eq{eq:Gm-r}. 
In fact the hyperangular-spin-isospin basis state $|K,m\rangle$ is
completely antisymmetric. However the sum over the permutations for
fixed values of $K$ produces linear dependent states that have to be
individuated and eliminated from the basis set~\cite{Viviani:1998,Viviani:2005,Dohet-Eraly:2019}.
This procedure could be delicate from a numerical point of view
as the number of $K$ increases. In such a case, one
needs a robust orthonormal procedure
capable to deal with the presence of large numerical
cancellations~\footnote{However, if one is successful in this step, at the end
  one can work with a basis of independent antisymmetrical states,
  whose number is noticeably less than the degenerancy of the full HH basis.}.
Attempts to use the HH basis
without symmetrization has been recently proposed~\cite{Gattobigio:2011}.
The idea here is to use the complete HH basis in which all symmetries
are represented to describe a particular state. The diagonalization of the 
Hamiltonian produces eigenvectors with well defined permutation properties
reflecting the symmetries in it. Different applications followed this procedure
for bosons as well as for fermions 
(see Refs.~\cite{Gattobigio:2009,Gattobigio:2011,
  Gattobigio:2011b,Deflorian:2013,Nannini:2018}). The advantage of
eliminating the orthonormalization of the states has to be balanced
by the fact that in this case one has to work with the
full basis of HH functions, whose degenerancy rapidly increases
with $K$ and the number of particles $A$.

Once the antisymmetric state $|K,m\rangle$ is constructed,
what is left is to obtain the unknown coefficients $c_{K;m}$
of the expansion. In order to do so, we apply the Rayleigh-Ritz
variational principle, which states that the quantity
$\langle \Psi_A |H-E|\Psi_A \rangle$ is stationary with respect to the variation
of any unknown coefficient. Here $H$ is the nuclear Hamiltonian and $E=-B$
the energy of the state, which, in the case of a bound state, is
negative and opposite to the binding energy $B$.

When differentiating respect to $c_{K;m}$ we obtain the following equation
\begin{equation}
  \sum_{K^\prime,m^\prime}\langle K, m | H | K^\prime,m^\prime\rangle c_{K^\prime;m^\prime}
  = E
  \sum_{K^\prime,m^\prime}\langle K, m | \unit | K^\prime,m^\prime\rangle c_{K^\prime;m^\prime}
  \label{eq:RR-eq} \ ,
\end{equation}
where the matrix elements of the Hamiltonian $H$ and of the identity
operator $\unit$
can be calculated with standard numerical techniques
(see Ref.~\cite{Kievsky:2008} for more details). Eq.~\eqref{eq:RR-eq}
represents a generalized eigenvalue-eigenvector problem, which can be
solved with a variety of numerical algorithms. Widely used within
the HH method is the Lanczos algorithm~\cite{Chen:1986},
since the HH basis can become quite large (up to
about 10,000 terms for $A=3$ and about one order of magnitude larger for $A=4$
are used in practice).

The results obtained solving Eq.~\eqref{eq:RR-eq}
for a variety
of nuclear interaction models will be presented in Sec.~\ref{sec:res}.

\subsubsection{The $A=3$ and $4$ scattering states}
\label{subsubsec:a34_scatt}
The HH method has been also applied to the scattering problem.
In particular, the method can study the elastic $N+Y\rightarrow N+Y$
process, where $N$ is a nucleon and $Y$ a bound system ($A_Y+1\equiv A = 3,4$),
both below and above the $Y$ nucleus breakup threshold. The extension
of the HH method to the full
breakup problem, i.e.\ for $A=3$ the process
$n+d\rightarrow n+n+p$, is currently underway
and will not be discussed here.

The wave function $\Psi^{LSJJ_z}_{NY}$ describing the $N+Y$
scattering state with incoming orbital angular momentum $L$,
channel spin ${\vec{S}}\equiv {\vec{\frac{1}{2}}}+{\vec{S}}_Y$,
parity $\pi=(-1)^L$, and
total angular momentum $J,J_z$, is written as
\beq\label{eq:psi_NY}
\Psi^{LSJJ_z}_{NY}=\Psi_C^{LSJJ_z}+\Psi_A^{LSJJ_z}
\ .
\eeq
Here we have introduced $\Psi_C^{LSJJ_z}$, which is the so-called
``core'' wave function, describing the system in the region
where all the particles are close to each other and their mutual
interaction is strong, and $\Psi_A^{LSJJ_z}$, which is the so-called
``asymptotic'' wave function, describing the relative motion
between nucleon $N$ and the nucleus $Y$ in the asymptotic region,
where the $N-Y$ interaction is negligible or 
reduces to the Coulomb interaction in the case of $N\equiv p$.
The core function $\Psi_C^{LSJJ_z}$ has to vanish at large $N-Y$
distance, and can be expanded in terms of the HH basis as for the bound
state. Therefore, using \Eq{eq:PsiA_final}, we can write
\beq\label{eq:psi_core}
\Psi_C^{LSJJ_z}=\sum_{K,m} c_{K;m} |K,m\rangle \ .
\eeq

The asymptotic wave function $\Psi_A^{LSJJ_z}$ is the solution of the
Schr\"odinger equation of the relative $N+Y$ motion. It
is written as a linear
combination of the following functions
\beq\label{eq:Omega_LSJJz}
\Omega^\lambda_{LSJJ_z}=\frac{\mathcal{C}}{\sqrt{N_p}}\sum_{p=1}^{N_p}
      [[\chi_{1/2}(N)\otimes\phi_{S_Y}(Y)]_S\otimes
        Y_L({\hat{\bm y}}_p)]_{J J_z}
      R^\lambda_L(y_p) \ .
\eeq
Here we have indicated with ${\mathcal{C}}$ a normalization factor (to be
explained below, see \Eq{eq:normOmega}).
The sum runs over the $N_p$ even permutations of the
$A$ nucleons necessary to antisymmetrize the function $\Omega^\lambda_{LSJJ_z}$,
$\chi_{1/2}(N)$ and $\phi_{S_Y}(Y)$ are the nucleon $N$ and nucleus $Y$
wave functions, respectively, and ${\bm y}_p$ is the relative distance
between
$N$ and the c.m. of nucleus $Y$ and is proportional to
$\bm{x}_{N-j+1}$ of \Eq{eq:defJac}.
Furthermore, $Y_L({\hat{\bm y}}_p)$ is the standard spherical harmonic
function, and the functions $R^\lambda_L(y_p)$ for $\lambda=R,I$ are
respectively the
regular and irregular solutions of the two-body $N+Y$ Schr\"odinger
equation without the nuclear interaction. They are explicitly
written as~\cite{Kievsky:2008,Marcucci:2009}
\beq
R^R_L(y_p)&=&\frac{1}{(2L+1)!! q^L C_L(\eta)}\frac{F_L(\eta,qy_p)}{qy_p} \ ,
 \label{eq:regf}\\
R^I_L(y_p)&=&(2L+1)!! q^{L+1} C_L(\eta) f(b,y_p) \frac{G_L(\eta,qy_p)}{qy_p} \ ,
 \label{eq:irregf}
\eeq
where $q$ is the modulus of the $N-Y$ relative momentum, such that the total
kinetic energy in the c.m. frame is $T_{c.m.}=q^2/2\mu$,
$\mu$ being the $N-Y$ reduced mass, $\eta=2Z_NZ_Y\mu e^2/q$
is the Coulomb parameter,
where $Z_N$ and $Z_Y$ are the charge numbers of $N$ and $Y$,
and $F_L(\eta,qy_p)$ and $G_L(\eta,qy_p)$ are
the regular and irregular Coulomb functions defined in the
standard way~\cite{Abramowitz}. The factor $C_L(\eta)$ is defined
in Ref.~\cite{Abramowitz} as
\beq\label{eq:CL}
C_L(\eta)=\frac{2^L{\rm e}^{-\frac{\pi\eta}{2}}|\Gamma(L+1+i\eta)|}{\Gamma(2L+2)}
\ .
\eeq
The factor $(2L+1)!! q^L C_L(\eta)$ has been introduced so that the functions
$R^R_L(y_p)$ and $R^I_L(y_p)$ have a finite limit for $q\rightarrow 0$.
Finally, the function $f(b,y_p)$ in \Eq{eq:irregf} is given by
\beq\label{eq:fby}
f(b,y_p)=[1-{\rm e}^{-by_p}]^{2L+1} \ ,
\eeq
so that the divergent behaviour of $G_L(\eta,qy_p)$ for small values of
$y_p$ is cured, and $R^I_L(y_p)$ is well-defined also in this limit.
The trial parameter $b$ is determined by requiring $f(b,y_p)\rightarrow 1$
for large values of $y_p$, leaving therefore unchanged the asymptotic
behaviour of the scattering wave function. A value of $b\sim 0.25$ fm$^{-1}$
has been found appropriate in all the considered cases. The non-Coulomb
case of \Eqs{eq:regf} and~\eqref{eq:irregf} is obtained if either $Z_N$
or $Z_Y=0$, so that the functions $F_L(\eta,qy_p)/(qy_p)$
and $G_L(\eta,qy_p)/(qy_p)$ are replaced by the regular and irregular
Riccati-Bessel functions $j_L(qy_p)$ and $n_L(qy_p)$ as
defined in Ref.~\cite{Abramowitz}, and the factor
$(2L+1)!! C_L(\eta)$ reduces to 1 for $\eta\rightarrow 0$~\cite{Abramowitz}.

With these definitions, $\Psi_A^{LSJJ_z}$  can be cast in the form
\beq\label{eq:psi_asymp}
  \Psi_A^{LSJJ_z}= \sum_{L^\prime S^\prime}
 \bigg[\delta_{L L^\prime} \delta_{S S^\prime} 
\Omega_{L^\prime S^\prime JJ_z}^R
  + {\mathcal R}^J_{LS,L^\prime S^\prime}(q)
     \Omega_{L^\prime S^\prime JJ_z}^I \bigg] \ ,
\eeq  
where the parameters ${\mathcal R}^J_{LS,L^\prime S^\prime}(q)$ give the
relative weight between the regular and irregular components 
of the wave function. These parameters can be written in terms of
the reactance matrix (${\mathcal K}$-matrix) elements
as~\cite{Kievsky:2008,Marcucci:2009}
\beq\label{eq:R-vs-K}
 {\mathcal K}^J_{LS,L^\prime S^\prime}(q)=
 (2L+1)!!(2L'+1)!!q^{L+L'+1}C_L(\eta)C_{L^\prime}(\eta)
 {\mathcal R}^J_{LS,L^\prime S^\prime}(q) \;\;\ .
\eeq
The ${\mathcal K}$-matrix, by definition, is such that its eigenvalues are
$\tan\delta_{LSJ}$, $\delta_{LSJ}$ being the phase shifts.
The sum over $L^\prime$ and $S^\prime$ in \Eq{eq:psi_asymp} is over all 
values compatible with a given $J$ and parity $\pi$,
and therefore the sum 
over $L^\prime$ is limited to include either even or odd values since
$(-1)^{L^\prime}=\pi$.

Using \Eqs{eq:psi_core} and~\eqref{eq:psi_asymp}, the full scattering wave
functions is written as
\beq\label{eq:psiNY-full}
\Psi^{LSJJ_z}_{NY}=\sum_{K,m} c_{K;m}|K,m\rangle+
\sum_{L^\prime S^\prime}
 \bigg[\delta_{L L^\prime} \delta_{S S^\prime} 
\Omega_{L^\prime S^\prime JJ_z}^R
  + {\mathcal R}^J_{LS,L^\prime S^\prime}(q)
  \Omega_{L^\prime S^\prime JJ_z}^I \bigg] \ ,
 \eeq
where the unknown quantities are the coefficients $c_{K;m}$ and
${\mathcal R}^J_{LS,L^\prime S^\prime}(q)$. In order to determine their values,
we use the Kohn variational principle~\cite{Kohn:1948}, 
which states that the functional
\beq\label{eq:kohn}
   [{\mathcal R}^J_{LS,L^\prime S^\prime}(q)]&=&
    {\mathcal R}^J_{LS,L^\prime S^\prime}(q)
     - \left \langle \Psi^{L^\prime S^\prime JJ_z }_{N-Y} \left |
         H-E \right |
        \Psi^{LSJJ_z}_{N-Y}\right \rangle \ , 
\eeq
has to be stationary with respect to variations of the trial parameters 
$c_{K;m}$ and
${\mathcal R}^J_{LS,L^\prime S^\prime}(q)$
in $\Psi^{LSJJ_z}_{NY}$. 
Here $E$ is the total energy of the system, and the normalization
coefficients ${\mathcal{C}}$
of the asymptotic functions $\Omega_{LSJJ_z}^\lambda$ in
\Eq{eq:Omega_LSJJz} are chosen so that
\beq\label{eq:normOmega}
   \langle \Omega^R_{LSJJ_z}| H-E | \Omega^I_{LSJJ_z} \rangle
  -\langle \Omega^I_{LSJJ_z}| H-E | \Omega^R_{LSJJ_z} \rangle =1 \ .
\eeq

The variation of the diagonal functionals of \Eq{eq:kohn} with respect
to the linear parameters $c_{K;m}$ leads to a system of linear
inhomogeneous equations,
\beq\label{eq:set1}
\sum_{K^\prime, m^\prime} \langle K,m| H-E |K^\prime,m^\prime
\rangle c_{K^\prime;m^\prime}^\lambda = 
     -D^{\lambda, LSJJ_z}_{K,m} \ ,
\eeq 
where the two terms $D^\lambda$ corresponding to
$\lambda\equiv R,I$ are defined as 
\beq\label{eq:dlm}
  D^{\lambda, LSJJ_z}_{K, m}= \langle K,m| H-E |
\Omega^\lambda_{LSJJ_z}\rangle \ .
\eeq
Therefore, two sets of the coefficients $c_{K;m}^\lambda$ are obtained,
depending on $\lambda\equiv R,I$,
and, consequently, we can introduce two core functions,
defined as
\beq\label{eq:psi_core_lambda}
\Psi_C^{LSJJ_z,\lambda}=\sum_{K,m}c_{K;m}^\lambda |K,m\rangle \ .
\eeq

The matrix elements ${\mathcal R}^J_{LS,L'S'}(q)$ are obtained 
varying the diagonal functionals of \Eq{eq:kohn} with respect to them. 
This leads to the following set of algebraic equations
\beq\label{eq:set2}
  \sum_{L'' S''} {\mathcal R}^J_{LS,L''S''}(q) X_{L'S',L''S''}= Y_{LS,L'S'} \ ,
\eeq
with the coefficients $X$ and $Y$ defined as
\beq
X_{LS,L'S'}&= \langle
\Omega^I_{LSJJ_z}+\Psi^{LSJJ_z,I}_C| H-E |\Omega^I_{L'S'JJ_z}\rangle \ ,
\nonumber \\
Y_{LS,L'S'}&=-\langle
\Omega^R_{LSJJ_z}+\Psi^{LSJJ_z,R}_C| H-E |\Omega^I_{L'S'JJ_z}\rangle \ .
\label{eq:xy}
\eeq

The solution of \Eq{eq:set2} provides a first-order estimate
of the matrix elements ${\mathcal R}^J_{LS,L'S'}(q)$. 
A second order estimate of ${\mathcal R}^J_{LS,L'S'}(q)$,
and consequently of ${\mathcal K}^J_{LS,L'S'}(q)$, is 
given by the quantities $[{\mathcal R}^J_{LS,L'S'}(q)]$, obtained by 
substituting in \Eq{eq:kohn} the
first order results of \Eqs{eq:set1} and~\eqref{eq:set2}.
Such second-order calculation provides then a symmetric 
${\mathcal K}$-matrix. This condition is not imposed {\it a priori},  
and therefore it is a useful test of the numerical accuracy
reached by the method.

The Kohn variational principle as explained so far is
particularly useful in the case of $q=0$ (zero-energy scattering).
For $q=0$ the scattering can occur only in the $L=0$ channel,
and the observables 
of interest are the scattering lengths. Within the 
present approach, they can be easily obtained from the relation
\beq\label{eq:scleng}
  ^{(2J+1)}a_{NY}=-\lim_{q\rightarrow 0}{\mathcal R}^J_{0J,0J}(q)\ .
\eeq

An alternative version of the Kohn variational principle
is the so-called complex Kohn variational principle for
the ${\mathcal S}$-matrix, quite convenient when 
$q\neq 0$ and especially above the $Y$ nucleus breakup threshold,
as explained in Ref.~\cite{Kievsky:1997}. In this case, the Kohn 
variational principle of \Eq{eq:kohn} becomes
\beq\label{eq:cmplxkohn}
   [{\mathcal S}^J_{LS,L^\prime S^\prime}(q)]=
    {\mathcal S}^J_{LS,L^\prime S^\prime}(q)
     + i \langle \Psi^{+,L^\prime S^\prime JJ_z}_{NY} |
         H-E |
        \Psi^{+,LSJJ_z}_{NY} \rangle \ ,
\eeq
where
\beq\label{eq:cmplxpsica}
    \Psi_{NY}^{+,LSJJ_z}=\Psi_C^{LSJJ_z}+\Psi_A^{+,LSJJ_z} \ ,
\eeq    
$\Psi_C^{LSJJ_z}$ being given in \Eq{eq:psi_core} and
\beq\label{eq:cmplxpsia}
\Psi_A^{+,LSJJ_z}&=&[\, i {\Omega}^R_{LSJJ_z} -
{\Omega}^I_{LSJJ_z}\,] \nonumber \\
&+&\sum_{L^\prime S^\prime}{\mathcal S}^J_{LS,L^\prime S^\prime}(q)
[\, i {\Omega}^R_{L^\prime S^\prime JJ_z} +
      {\Omega}^I_{L^\prime S^\prime JJ_z}\, ]
\ . 
\eeq
The functions ${\Omega}^\lambda_{LSJJ_z}$ have been given in \Eq{eq:Omega_LSJJz}.
Note that, with the above definition, the reactance ${\mathcal K}$-matrix 
elements can be related to the ${\mathcal S}$-matrix elements as
\beq\label{eq:ksmt}
{\mathcal K}^J_{LS,L^\prime S^\prime}(q)=(-i)
[{\mathcal S}^J_{LS,L^\prime S^\prime}(q) - \delta_{LL^\prime}\delta_{S S^\prime}]
\,
[{\mathcal S}^J_{LS,L^\prime S^\prime}(q) + \delta_{LL^\prime}\delta_{S S^\prime}]^{-1}
\ . 
\eeq
The differentiation of the complex Kohn variational principle of
\Eq{eq:cmplxkohn} leads to a set of equations for $c_{K;m}$
and ${\mathcal S}^J_{LS,L^\prime S^\prime}(q)$ similar to those
given in \Eqs{eq:set1} and~\eqref{eq:set2}, where now
$\lambda$ stands for $\lambda=+,-$.

We conclude this section with the following remarks: (i) the calculation
of the matrix elements involving $\Psi_C^{LSJJ_z}$ can be performed
with the HH expansion either in coordinate- or in momentum-space, depending
on what is more convenient. Therefore, regarding this part, we can
apply the method with any potential model, both local
or non local. (ii) Some difficulties arise with the calculations
of the potential matrix elements which involve 
$\Omega^\lambda_{LSJJ_z}$, i.e. 
$\langle K,m |V|\Omega^\lambda_{LSJJ_z}\rangle$ present in 
\Eq{eq:dlm}, and 
$\langle\Omega^{\lambda '}_{L'S'JJ_z}+\Psi_C^{L'S'JJ_Z,\lambda'} |V|
\Omega^\lambda_{LSJJ_z}\rangle$ of \Eq{eq:xy}, 
with $\lambda, \lambda '=R$, $I$. All technical details
can be found in Ref.~\cite{Marcucci:2009}. We note here
that these difficulties have been overcome in the case
of all local potentials and non-local projecting potentials,
like the recent chiral and $V_{low-k}$ potential models.
On the other hand, some difficulties remain for the
non-local meson-theoretic CDBonn potential model. 
(iii) The three-nucleon interaction models which at the moment
have been implemented with the HH method are only the local
ones, like the Urbana IX potential (UIX) of Ref.~\cite{Pudliner:1995}
and the N2LO model of Ref.~\cite{Navratil:2007}. The models used so-far,
besides being local, 
have a well defined operatorial structure. In this case, 
the projection procedure as used for the two-nucleon interaction
is not needed and the approach
follows well-established footsteps, as explained
in Refs.~\cite{Kievsky:1994,Kievsky:1996}.

%
%

\section{Selected results}
\label{sec:res}

We present in this section selected results obtained with the
HH method described above. The method has been applied
widely since many years, and therefore a selection is mandatory.
We have followed these criteria: (i) we focus on the results
obtained after 2008, year of the publication
of the review of Ref.~\cite{Kievsky:2008} on the same method.
(ii) We restrict ourselves
to the potential models, mostly discussed in the Research Topic of
which this contribution is part. They are the most
widely used models. (iii) We concentrate on the results obtained for the
$A=3,4$ elastic scattering observables, but we present briefly also
the corresponding bound state results.

The aim of this section is
twofold: first of all we wish to show the effectiveness of the HH method
for few-nucleon systems; secondly, we want to emphasize that
the HH method, as well as any {\it ab-initio} method, is an essential
tool for testing and eventually improving nuclear interaction models.

The potentials which will appear in the following subsections
include both two- and three-nucleon interactions. They are
the phenomenological two-nucleon interaction
Argonne $v_{18}$ (AV18)~\cite{Wiringa:1995},
augmented by the three-nucleon Urbana IX (UIX) model~\cite{Pudliner:1995},
the meson-theoretic CDBonn potential~\cite{Machleidt:2001} (CDB),
together with the three-nucleon Tucson-Melbourne~\cite{Coon:1980,Coon:2001} (TM) model,
and the $V_{low-k}$ potential~\cite{Bogner:2006},
obtained from the AV18 with $\Lambda=2.2$~fm$^{-1}$,
so that the triton binding energy is reproduced.
We consider in addition also 
chiral potentials, in particular the two-nucleon interaction
models of the Idaho group of Ref.~\cite{Entem:2003},
obtained at next-to-next-to-next-to-leading order (N3LO),
and here labeled with N3LO-I, and the 
more recent models derived by the same group in Ref.~\cite{Entem:2017},
here labeled according to the chiral order, i.e.\
from leading order (LO)
up to N4LO. All these two-nucleon models have been augmented with a (local)
three-nucleon interaction derived up to N2LO as in 
Ref.~\cite{Navratil:2007}.
The momentum-cutoff value is set equal to $\Lambda=500$ MeV,
unless differently specified.
Note that the low-energy constants (LECs) $c_D$ and $c_E$
are those 
of Ref.~\cite{Navratil:2007} when the N2LO
three-nucleon interaction is used in conjunction with the N3LO-I
two-nucleon potential, while the LECs are those of
Ref.~\cite{Marcucci:2019} when the N2LO three-nucleon
interaction is used in
conjunction with the N2LO, N3LO and N4LO two-nucleon interactions
of Ref.~\cite{Entem:2017} (no three-nucleon interaction is present
at lower chiral order).
Finally, we will present results obtained also with the
minimally non-local chiral potentials of the Norfolk
group, as derived in Ref.~\cite{Piarulli:2015} for the two-nucleon,
and in Refs.~\cite{Piarulli:2018,Baroni:2018}
for the thee-nucleon interaction.
The two-nucleon models are labeled NVIa, NVIIa, NVIb and NVIIb depending on the
cutoff value and the maximum laboratory energy of the considered $NN$
database. When the three-nucleon interaction are included, we will
refer to NV2+3/Ia, NV2+3/IIa, and so on, corresponding
to the fitting procedure of Ref.~\cite{Piarulli:2018}, and
NV2+3/Ia*, NV2+3/IIa*, and so on, corresponding
to the fitting procedure of Ref.~\cite{Baroni:2018}.
We discuss in more details these fitting procedures below,
and we refer the reader to the original references,
or to the contributions present in this Research Topic.
To be noticed that when the HH method is used to study the
bound states, the local AV18, AV18/UIX, NV, and NV2+3 potentials
have been all augmented by the full electromagnetic interaction, which
includes corrections up to $\alpha^2$ ($\alpha$ is the fine-structure
constant). On the other hand, the non-local CDB, CDB/TM, and all the non-local
chiral potentials retain only the point-Coulomb interaction.
The point-Coulomb interaction,
and not the full electromagnetic one, is also used for the
scattering states results presented below.

\subsection{$A=3,4$ bound states}
\label{subsec:bound}

The results for the trinucleon and $^4$He binding energies,
obtained using 
all the above mentioned potentials, are given in Table~\ref{tab:bound}.
To be noticed that
in many cases, the experimental trinucleon binding energy is used for the LECs
fitting procedure. When this occurs, the
corresponding HH results is underlined in the table.
We briefly outline the fitting procedure
in order to better understand the results,
and we refer to Refs.~\cite{Marcucci:2012,Piarulli:2018,Baroni:2018}
for more details.
The $^3$H and $^3$He ground state wave functions are calculated using
a given two- and three-nucleon potential, and the
corresponding LECs $c_D$ and $c_E$ are determined by fitting the
$A=3$ experimental binding energies, corrected for a small contribution
($+7$ keV in $^3$H and $-7$ keV in $^3$He), due to the $n-p$ mass
difference~\cite{Nogga:2003}, since in the present
HH method this effect is neglected.
This procedure generates two trajectories, one for $^3$H and
one for $^3$He, in the $\{c_D,c_E\}$ plane, so
that each point of the trajectory corresponds to the correct binding
energy. The two trajectories are typically extremely close to each other
and the average can be safely considered, since the points of the
average trajectory
typically lead to $A=3$ binding energies within 10 keV of the
experimental values. 
A second observable is needed in the fitting procedure.
In Ref.~\cite{Piarulli:2018}  the $n-d$ doublet scattering
length $^2a_{nd}$ has been used, which leads in the $\{c_D,c_E\}$ plane to another trajectory,
which is very close to the one corresponding to the $^3$H binding energy,
but not exactly overlapping. This is a well-known fact, that the
$^3$H binding energy and $^2a_{nd}$ are correlated
observables. However, it is possible to find an intersection point of the two
trajectories, which allows to determine the LECs.
This procedure has been used for the NV2+3/Ia, NV2+3/Ib, NV2+3/IIa,
and NV2+3/IIb potential models.
The corresponding $\{c_D,c_E\}$ values, as given in Table I
    of Ref.~\cite{Piarulli:2018}, are
    $\{3.666, -1.638\}$, $\{-2.061,-0.982\}$, $\{1.278,-1.029\}$,
    $\{-4.480,-0.412\}$, respectively.
Alternatively we can choose as the second observable
the Gamow-Teller matrix element of tritium $\beta$-decay, to take
advantage of the fact that the LEC $c_D$ enters also in the
two-nucleon axial current operator at
N2LO~\cite{Marcucci:2012,Baroni:2018,Gardestig:2006,Gazit:2009}.
This second procedure has been used for the N2LO/N2LO, N3LO/N2LO and
N4LO/N2LO potentials of Ref.~\cite{Marcucci:2019},
and the NV2+3/Ia*, NV2+3/Ib*, NV2+3/IIa*,
and NV2+3/IIb* potential models of Ref.~\cite{Baroni:2018}.
In this last case, we report the corresponding $\{c_D,c_E\}$ values
for completeness, which are $\{-0.635,-0.090\}$, $\{-4.710,0.550\}$,
$\{-0.610,-0.350\}$, $\{-5.250,0.050\}$,
respectively.

We can now proceed with some comments regarding the binding energies
results of Table~\ref{tab:bound}. (i) The large variety of models
for the nuclear interaction which the HH method can handle is
an indication of how strong and reliable this method has become.
Furthermore, we should mention that the theoretical uncertainty
is of 1 keV (10 keV) for the $A=3$ ($^4$He) binding energies.
The HH method is therefore extremely accurate.
Furthermore, all the HH results are in very good agreement with the
values reported in the literature, when available.
(ii) In order to reproduce the experimental binding energies
 the inclusion of three-nucleon force is essential.
In all cases, the triton binding energy is well reproduced, within few
keV. On the other hand, the
$^4$He binding energies can differ from the experimental
value of even up to 700 keV (in the CDB/TM case).
(iii) In the case of the NV2+3 potential models, when
the observables used to fit the LECs
are the triton binding energy and $^2a_{nd}$,
we notice a systematic overestimation of the $^3$He
binding energy.
(iv) All the results for the $A=3$ ($A=4$) binding energies
obtained with any model for the two- and three-nucleon interaction
are within 10 (400) keV from the experimental values.
Therefore we can conclude that any of the constructed model is essentially 
able to reproduce these very light nuclei.

\begin{table}[htb]
\caption{\label{tab:bound}
  The binding energies in MeV for $^3$H, $^3$He and $^4$He,
  calculated with the HH technique using different Hamiltonian models.
  The underlined values are used in the LECs fitting procedure.
  In the last row, we show the $^3$H ($^3$He) experimental binding energy of
  8.482 MeV (7.718 MeV), lowered (increased) by 7 keV in order to take into
  account the $n-p$ mass difference. See text for more details.
  All the results presented here are in very good agreement with the
  values reported in the literature.}
\begin{center}
\begin{tabular}{l|ccc}
\hline
Interaction & $^3$H & $^3$He & $^4$He \\
\hline
AV18 & 7.624 & 6.925 & 24.21 \\
AV18/UIX & 8.479 & 7.750 & 28.46 \\
CDB & 7.998 & 7.263 & 26.13 \\
CDB/TM & 8.474 & 7.720 & 29.00 \\
\hline
N3LO-I & 7.854 & 7.128 & 25.38 \\
N3LO-I/N2LO & 8.474 & 7.733 & 28.36 \\
\hline
LO & 11.091 & 10.409 & 40.09 \\
NLO & 8.307 & 7.597 & 27.55 \\
N2LO & 8.206 & 7.460 & 27.23 \\
N3LO & 8.092 & 7.343 & 26.68\\
N4LO & 8.080 & 7.337 & 26.58 \\
N2LO/N2LO & {\underline{8.474}} & {\underline{7.729}}& 27.92 \\
N3LO/N2LO & {\underline{8.477}} & {\underline{7.728}}& 27.97 \\
N4LO/N2LO & {\underline{8.477}} & {\underline{7.728}}& 28.15 \\
\hline
NVIa & 7.818 & 7.090 & 25.15 \\
NVIIa & 7.949 & 7.213 & 25.80 \\
NVIb & 7.599 & 6.885 & 23.96 \\
NVIIb & 7.866 & 7.133 & 25.28 \\
\hline
NV2+3/Ia & {\underline{8.475}}& 7.735 & 28.33 \\
NV2+3/IIa & {\underline{8.475}} & 7.730 & 28.16 \\
NV2+3/Ib & {\underline{8.475}} & 7.737 & 28.30 \\
NV2+3/IIb & {\underline{8.475}}& 7.727 & 28.15 \\
\hline
NV2+3/Ia* & {\underline{8.477}}& {\underline{7.727}}& 28.30 \\
NV2+3/IIa* & {\underline{8.474}}& {\underline{7.725}}& 28.18 \\
NV2+3/Ib* & {\underline{8.469}}& {\underline{7.724}}& 28.21 \\
NV2+3/IIb* & {\underline{8.474}}& {\underline{7.724}}& 28.11 \\
\hline
Experiment & 8.475 & 7.725 & 28.30 \\
\hline
\end{tabular}
\end{center}
\end{table}

\subsection{$N-d$ scattering}
\label{subsec:nd-scattering}

One of the remarkable features of the HH method resides in its capability
of dealing with local as well as with non-local potentials, formulated in
either coordinate or momentum space, not only for the
bound states, as we have seen above, but also for $N-d$
scattering state observables.
This has been demonstrated in
Ref.~\cite{Marcucci:2009}, in which the local AV18 and the non-local
chiral N3LO-I potential models were used to
calculate the $N-d$ elastic scattering observables
below the deuteron breakup threshold.
Here we present results with a subset of all the
potential models mentioned above, and in particular
with the AV18, AV18/UIX, the N3LO-I, N3LO-I/N2LO, and some of the NV and NV2+3
models.
A further class of nuclear interactions that has been tested
using the HH method is represented by
the so-called $V_{low-k}$ potential
obtained from the AV18 with $\Lambda=2.2$~fm$^{-1}$,
so chosen to reproduce the triton binding energy when the complete
electromagnetic interaction is used~\cite{Bogner:2006}.
We do not report here 
detailed investigations on the
convergence of the HH expansion, but we can mentioned that
this convergence is faster for the
non-local potentials as compared to the local ones, due to the much
softer behaviour at small distances. For instance, for $N-d$
elastic scattering in the channel $J^\pi=1/2^+$,
the HH basis can be of the order of $12000$ ($7000$) elements with the
NV (N3LO-I) potential to get convergence.
\begin{table}[htb]
\caption{\label{tab:sl}
$n-d$ and $p-d$ doublet and quartet scattering lengths in fm
  calculated with the HH technique using different Hamiltonian models.
  The experimental value for $^2a_{nd}$ is from Ref.~\cite{Schoen:2003},
  while that for $^4a_{nd}$  is from Ref.~\cite{Dilg:1971}.}
\begin{center}
\begin{tabular}{l|cccc}
\hline
Interaction & $^2a_{nd}$ & $^4a_{nd}$ & $^2a_{pd}$ & $^4a_{pd}$ \\
\hline
AV18 & 1.275 & 6.325 & 1.185 & 13.588 \\
AV18/UIX & 0.610 & 6.323 & -0.035 & 13.588 \\
\hline
$V_{low-k}$ & 0.572 & 6.321 & -0.001 & 13.571 \\
\hline
N3LO-I & 1.099 & 6.342 & 0.876 & 13.646 \\
N3LO-I/N2LO & 0.675 & 6.342 & 0.072 & 13.647 \\
\hline
NVIa & 1.119 & 6.326 & 0.959 & 13.596 \\
NVIb & 1.307 & 6.327 & 1.294 & 13.597 \\
NV2+3/Ia* & 0.638 & 6.326 & 0.070 & 13.596 \\
NV2+3/Ib* & 0.650 & 6.327 & 0.070 & 13.597 \\
\hline
Experiment & 0.645$\pm 0.003\pm 0.007$ & 6.35$\pm 0.02$ & -- & -- \\
\hline
\end{tabular}
\end{center}
\end{table}

We first consider the converged results for the $n-d$ and $p-d$
doublet and quartet scattering
lengths, which are given in Table~\ref{tab:sl}, together with the very precise
experimental result from Ref.~\cite{Schoen:2003} for $^2a_{nd}$,
and the older experimental results from
Ref.~\cite{Dilg:1971} for $^4a_{nd}$. No experimental
data are available for $^{2}a_{pd}$ and $^4a_{pd}$.
All the results are obtained using the pure Coulomb electromagnetic
interaction. When the full electromagnetic interaction is used,
$^4a_{nd}$ remains practically unchanged, while $^2a_{nd}$ becomes smaller.
For the NVIa and NVIb potentials, for instance,
$^2a_{nd}=1.103$ fm and $1.293$ fm, respectively, with the
full electromagnetic interaction.
As it is clear from inspection of Table~\ref{tab:sl}, while $^4a_{nd}$ is very
little model-dependent and in good agreement with experiment,
the same is not true for $^2a_{nd}$.
In particular, the inclusion of a three-nucleon force
appears necessary to bring
the results closer to the experimental datum. However, although the chiral
potentials give slightly better results, none of the considered models agrees
with the experiment. The disagreement is more pronounced for the
$V_{low-k}$ interaction, showing that this observable cannot be
simply reproduced by increasing the attraction
of the two-nucleon interaction, as is done in this case by
choosing the right value for
$\Lambda$ to describe the triton; instead, a subtle balance
between attraction and repulsion in the three-nucleon system has to be
reached. Indeed, being the zero-energy $n-d$ scattering state orthogonal
to the triton, the associated wave function presents a node in the relative
distance, whose precise position, which is related to the scattering length,
depends on the interplay between attraction and repulsion.

\begin{figure}[htb]
\begin{center}
\includegraphics[scale=0.6]{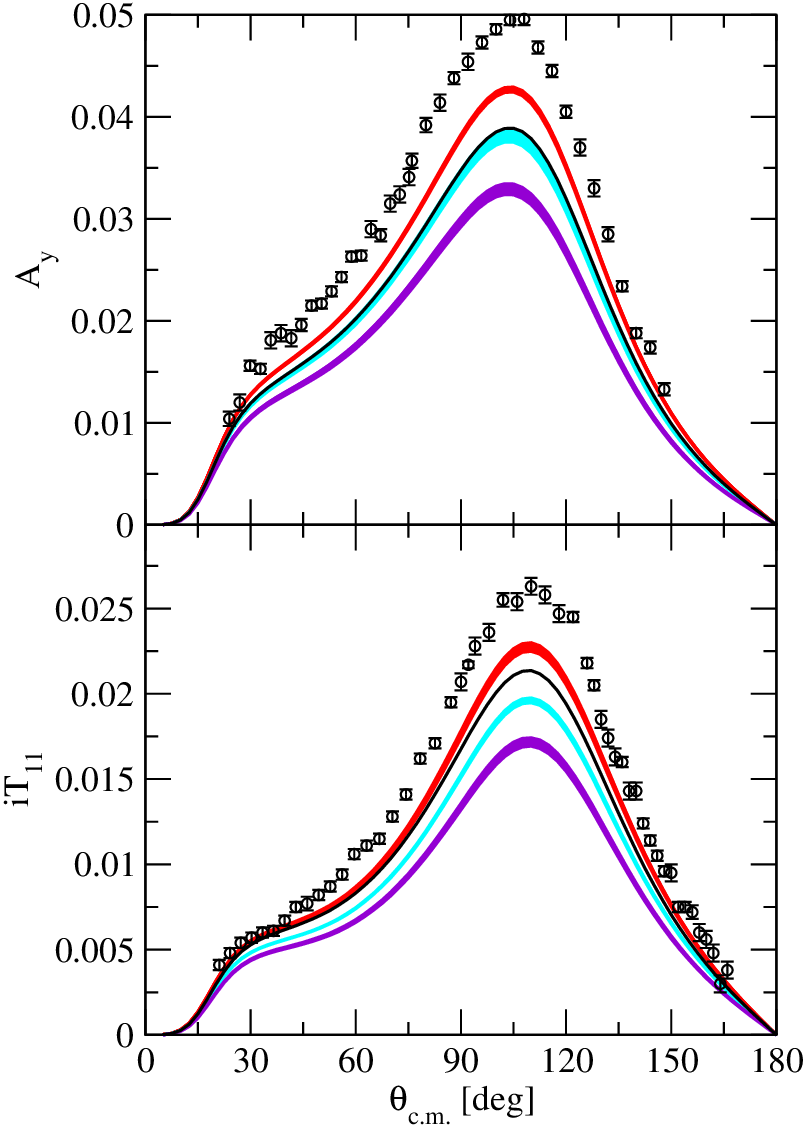}
\caption{(Color online) The vector analyzing powers $A_y$ and $iT_{11}$
  at center-of-mass energy $E_{c.m.}=2$ MeV, using models
  in the AV18/TM class (cyan bands), AV18/UIX (violet bands)
  and AV18/N2LO (red bands).
  The predictions of the original AV18/UIX model (solid lines)
  and the experimental points from Ref.~\protect\cite{Shimizu:1995}
  are also shown.}
\label{fig:ayfig}
\end{center}
\end{figure}

With the purpose of investigating the capability of some widely used models
of three-nucleon interaction
to reproduce $^2a_{nd}$, a sensitivity study was conducted in
Ref.~\cite{Kievsky:2010}  taking the AV18 as the reference two-nucleon
interaction. Three different models of the three-nucleon interactions
were considered: the UIX, the TM and the chiral N2LO of
Ref.~\cite{Navratil:2007}.
Their parameters were adjusted, constraining  them to reproduce simultaneously
$^2a_{nd}$ and the triton binding energy, and the resulting value
for the $^4$He binding energy was calculated.
For the UIX model, a reasonable description of these three observables was
possible, at the cost of a sizable increase of the repulsive term, as
compared to the original parameterization. A similar conclusion held for the
TM model, where a repulsive short-range term was found to be necessary.
Finally, for the N2LO three-nucleon interaction,
the relative importance of the parameters
involving the $P$-wave pion rescattering had to be changed. This is not
surprising, due to the mismatch between the physics underlying the adopted
models for two- and three-nucleon interactions. Also in this case, a
repulsive short-range interaction was preferred.
Then, a set of polarization observables on elastic $p-d$
scattering were computed using the AV18 augmented by the modified versions of the three-nucleon interactions models as described above. These led to three classes of interaction models.
As an interesting result, all models within a given class
led to very similar
predictions, but for some observables, namely the proton $A_y$ and the
deuteron $i T_{11}$, these predictions were different
from class to class, and all in disagreement with the data.
This is shown in Fig.~\ref{fig:ayfig}.
Since the three classes of models  mostly  differ in their short-distance
behavior,  it follows that an improvement in this component of the
three-nucleon interaction is
needed to explain the data. Indeed, no sensible improvement was obtained as
compared to the original AV18/UIX model.

In order to be more quantitative, as to the accuracy of the existing models
of two- and three-nucleon interaction,
we show in Table~\ref{tab:chi2} the $\chi^2/$datum for all
$p-d$ elastic scattering observables
at different center-of-mass energies, as obtained with the AV18 and
N3LO-I two-body interactions, without or with
the inclusion also of the UIX and N2LO
three-nucleon interaction models~\cite{Marcucci:2009}.
\begin{table}[htb]
\caption{\label{tab:chi2}
$\chi^2$/datum of the $p-d$ elastic scattering observables
at center-of-mass energies $E_{c.m.}=0.666, 1.33, 1.66$ and 2.0 MeV, 
calculated with the N3LO-I or AV18
two-nucleon only, and the N3LO-I/N2LO or AV18/UIX two- and
three-nucleon Hamiltonian models. 
The different number $N$ of experimental data is also indicated.
The data are from Refs.~\protect\cite{Kievsky:2000,Wood:2001}
at $E_{c.m.}=0.666$ MeV, 
and from Ref.~\cite{Shimizu:1995} at $E_{c.m.}=1.33, 1.66$ and 2.0 MeV.}
\begin{footnotesize}
\begin{center}
\begin{tabular}{c|ccccc|c|ccccc|ccccc}
\hline
& \multicolumn{5}{c|}{0.666 MeV} & 1.33 MeV 
& \multicolumn{5}{c|}{1.66 MeV} & \multicolumn{5}{c}{2.0 MeV} \\
& $A_y$ & $iT_{11}$ & $T_{20}$ & $T_{21}$ & $T_{22}$ 
& $A_y$ 
& $A_y$ & $iT_{11}$ & $T_{20}$ & $T_{21}$ & $T_{22}$ 
& $A_y$ & $iT_{11}$ & $T_{20}$ & $T_{21}$ & $T_{22}$ \\
\hline
$N$ & 7 & 8 & 24 & 24 & 24 
& 38 
& 44 & 50 & 50 & 50 & 50 
& 38 & 51 & 51 & 51 & 51   \\
\hline
AV18 & 283.3 & 113.4 & 6.9 & 4.7 & 2.8
& 186.0
& 267.6 & 121.3 & 1.9 & 3.2 & 6.6
& 237.1 & 148.8 & 3.7 & 5.1 & 12.5 \\
AV18/UIX & 205.2 & 67.0 & 3.2 & 3.5 & 1.1
& 112.4
& 264.7 & 110.1 & 4.2 & 7.2 & 2.1
& 202.4 & 115.0 & 6.4 & 14.3 & 2.2 \\
N3LO-I & 197.7 & 68.7 & 4.0 & 2.6 &  1.5 
& 108.4 
& 227.9 & 92.6 & 1.0 & 2.2 &  2.7 
& 186.0 & 108.3 & 1.9 & 2.8 &  4.4 \\
N3LO-I/N2LO & 139.9 & 49.5 & 2.7 & 2.5 & 0.9 
& 70.0
& 159.4 & 84.3 & 2.1 & 4.0 & 2.8 
& 114.0 & 85.8 & 3.6 & 8.3 & 1.6 \\
\hline
\end{tabular}
\end{center}
\end{footnotesize}
\end{table}
It is clear that all considered models fail to give a satisfactory
description of all polarization observables, especially for $A_y$ and
$i T_{11}$. From the previous discussion, there are strong hints that the
improvement should come from a more accurate modeling of the short distance
structure of the three-nucleon interaction. Therefore,
in Ref.~\cite{Girlanda:2011} all the possible short-distance (contact)
structures for the three-nucleon interaction
have been classified up to the subleading order of
a systematic low-energy expansion. It has been found that the
short-distance component of the
three-nucleon interaction
can be parametrized by ten LECs, denoted by
$E_i$ with $i=1,...,10$.
The corresponding three-nucleon potential
in configuration space can be written as
\begin{eqnarray} \label{eq:vcontact2}
  V_{3N{\mathrm{cont}}}&=&\sum_{i\neq j\neq k} E_0 Z_0(r_{ij};\Lambda)
  Z_0(r_{ik};\Lambda) \nonumber \\
  &&+
  (E_1 + E_2 {\bm \tau}_i \cdot {\bm \tau}_j + E_3 {\bm \sigma}_i \cdot
  {\bm \sigma}_j + E_4 {\bm \tau}_i \cdot {\bm \tau}_j
  {\bm \sigma}_i \cdot {\bm \sigma}_j)  \left[
    Z_0^{\prime\prime}(r_{ij};\Lambda) + 2 \frac{Z_0^\prime(r_{ij};\Lambda)}{r_{ij}}
    \right]
  Z_0(r_{ik};\Lambda)  \nonumber \\
  && + (E_5 +E_6 {\bm \tau}_i\cdot{\bm \tau}_j) S_{ij} \left[
    Z_0^{\prime\prime}(r_{ij};\Lambda) - \frac{Z_0^\prime(r_{ij};\Lambda)}{r_{ij}}
    \right]
  Z_0(r_{ik};\Lambda) \nonumber \\
  && + (E_7 + E_8 {\bm \tau}_i\cdot{\bm \tau}_k) ( {\bf L}\cdot {\bm S})_{ij}
  \frac{Z_0^\prime(r_{ij};\Lambda)}{r_{ij}} Z_0(r_{ik};\Lambda) \nonumber \\
  && + (E_9 + E_{10} {\bm \tau}_j \cdot {\bm \tau}_k)
       {\bm \sigma}_j \cdot \hat {\bf r}_{ij}
       {\bm \sigma}_k \cdot \hat {\bf r}_{ik} Z_0^\prime(r_{ij};\Lambda)
       Z_0^\prime(r_{ik};\Lambda) \ ,
\end{eqnarray}
where ${\bm \sigma}_i$ (${\bm \tau}_i$) are the Pauli spin (isospin)
matrices of particle $i$, ${\bf r}_{ij}$ is the relative distance
between particles $i$ and $j$, and
$S_{ij}$ and $ ( {\bf L}\cdot {\bm S})_{ij}$ are, respectively, the
tensor and spin-orbit operators. The profile
functions $Z_0(r;\Lambda)$ are  written as
\begin{equation}
  Z_0(r;\Lambda)=\int \frac{d {\bf k}}{(2 \pi)^3}
  {\mathrm{e}}^{i {\bf k}\cdot {\bf r}} F({\bf k}^2;\Lambda) \ ,
  \label{eq:z0}
\end{equation}
with $F({\bf k}^2; \Lambda)$  a suitable cutoff function which suppresses
the momentum transfers ${\bf k}$ below a given short-distance cutoff $\Lambda$.
In Eq.~(\ref{eq:vcontact2}),
the basis of operators
has been chosen so that most terms in the potential can be viewed as an
ordinary interaction of particles $ij$ with a further dependence on the
coordinate of the third particle $k$. 
In Ref.~\cite{Girlanda:2018}, elastic $p-d$ scattering data at
$E_{c.m.}=2$ MeV center-of-mass energy have been used to fit the
$E_i$ LECs, when the subleading
three-nucleon interaction given in Eq.~(\ref{eq:vcontact2}) is considered
in addition
to the AV18/UIX interaction.
Also $^2a_{nd}$ and $^4a_{nd}$ and the triton binding energy have been
included in the fit.
\begin{table}[htb]
\begin{center}
  \begin{tabular}{ccccc}
    \hline
$\Lambda$ (MeV)    & 200&300 &400 &500
    \\
    \hline
    $\chi^2$/datum & 2.0 & 2.0 & 2.1 & 2.1 \\
    $e_0$ & -0.074 & -0.037 & 0.053 & 0.451 \\
    $e_5$ & -0.212 & -0.248 & -0.403 & -0.799 \\
    $e_7$ & 1.104 & 1.195 & 1.686 & 2.598 \\
\hline
$\langle {\mathrm{AV18}}\rangle$ (MeV)& -7.353 & -7.373 & -7.394 & -7.343 \\
$\langle {\mathrm{UIX}}\rangle$ (MeV)&-1.118 & -1.095 & -1.058 & -1.031 \\
$\langle E_0\rangle$ (MeV) &-0.057  & -0.069 & 0.125 & 0.841 \\
$\langle E_5 O_5\rangle$ (MeV) &-0.032 & -0.182 & -0.609 & -1.553 \\
$\langle E_7 O_7\rangle$ (MeV) &0.079 & 0.237 & 0.454 & 0.605 \\
\hline
$^2a_{nd}$ (fm) & 0.611 & 0.618 & 0.626 &0.638 \\ 
    $^4a_{nd}$ (fm) & 6.32 & 6.32 &6.32 &6.32\\
    \hline
  \end{tabular}
\end{center}
\caption{$\chi^2$/datum of the two-parameter fit obtained neglecting
  in Eq.~(\ref{eq:vcontact2}) all the subleading operators
  except the leading contact term proportional
  to the LEC $E_0$, and the tensor
  and spin-orbit operators, indicated with $O_5$ and $O_7$
  respectively, proportional to the LECs $E_5$ and
  $E_7$, considered on top of the AV18/UIX potential model.
  The LECs $e_0,e_5,e_7$ are defined in terms of $E_0,E_5,E_7$
  as $E_0=e_0/(F^4_\pi\Lambda)$, $E_i=e_i(F^4_\pi\Lambda^3)$, $i=5,7$,
  $F_\pi=92.4$ MeV being the pion decay constant, so that
  $e_0\sim e_i\sim O(1)$ if natural.
  Also shown are the mean values in the triton state of the
  one- plus two-body Hamiltonian
  (labeled as $\langle \mathrm{AV18}\rangle$), of the UIX three-body
  potential (labeled as $\langle \mathrm{UIX}\rangle$), and of
  individual contributions from the short-distance three-body potential.
  The calculated values of $^2a_{nd}$ and $^4a_{nd}$ are also given.}
\label{tab:2par}
\end{table}

\begin{figure}[htb]
  \centerline{\includegraphics[scale=0.8]{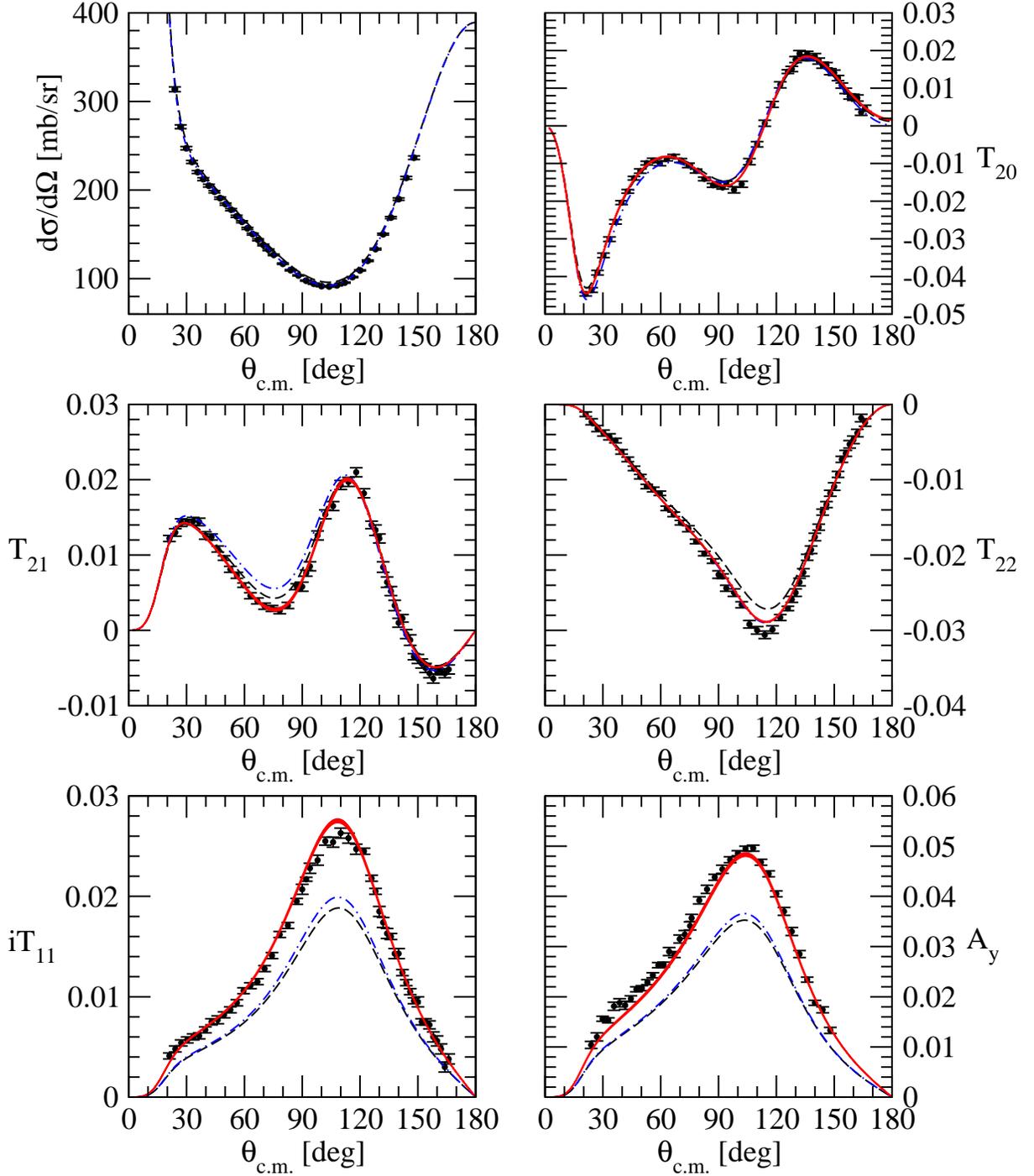}}
  \caption{(Color online)
    Curves obtained including only the tensor and spin-orbit
    subleading contact operator on the top of the AV18/UIX  interaction,
    fitted to a set of cross section and polarization observables in $p-d$
    elastic scattering at 2~MeV center-of-mass energy~\cite{Shimizu:1995}, 
    for $\Lambda=200-500$~MeV (red bands), are compared to the purely
    two-body AV18 interaction (dashed black lines) and to the AV18/UIX 
    two- and three-nucleon interaction (dashed-dotted blue lines).}
    \label{fig:fit57}
\end{figure}

The results of Ref.~\cite{Girlanda:2018} can be summarized as follows.
First of all, we noticed that the operators which play a leading role
in reducing the large $\chi^2$/datum of Table~\ref{tab:chi2} are
the spin-orbit and tensor interactions, which depend on the LECs
$E_5$ and $E_7$. We present in Table~\ref{tab:2par} the results
of a fit where only the terms proportional to $E_0$,
$E_5$ and $E_7$ are kept.
The LEC $E_0$ is used to fix the triton binding energy. Then
the experimental data for the doublet and quartet $n-d$ scattering lengths
of Refs.~\cite{Schoen:2003} and~\cite{Dilg:1971}, and those
of several $p-d$ scattering observables at 2 MeV center-of-mass
energy
of Ref.~\cite{Shimizu:1995} are used for the determinations of the LECs.
As it is shown in Table~\ref{tab:2par}, the $\chi^2$/datum is drastically
reduced to $\sim 2$ 
for the short distance cutoff $\Lambda$ of Eq.~(\ref{eq:z0})
between 200 and 500~MeV.
More sophisticated fits, including all the involved LECs, lead to only
slightly better $\chi^2/$datum $\sim 1.6$.
In Fig.~\ref{fig:fit57} we show the corresponding fitted curves compared to
the AV18 and AV18/UIX predictions. It is clear that a very accurate description
can be obtained with only the spin-orbit and tensor subleading operators.
We also note that the values of the LECs $e_0,e_5,e_7$, defined in terms of 
$E_0,E_5,E_7$ as $E_0=e_0/(F^4_\pi\Lambda)$, $E_i=e_i(F^4_\pi\Lambda^3)$, $i=5,7$,
$F_\pi=92.4$ MeV being the pion decay constant, are of order 1
as expected.

\begin{figure}[htb]
  \centerline{\includegraphics[scale=0.8]{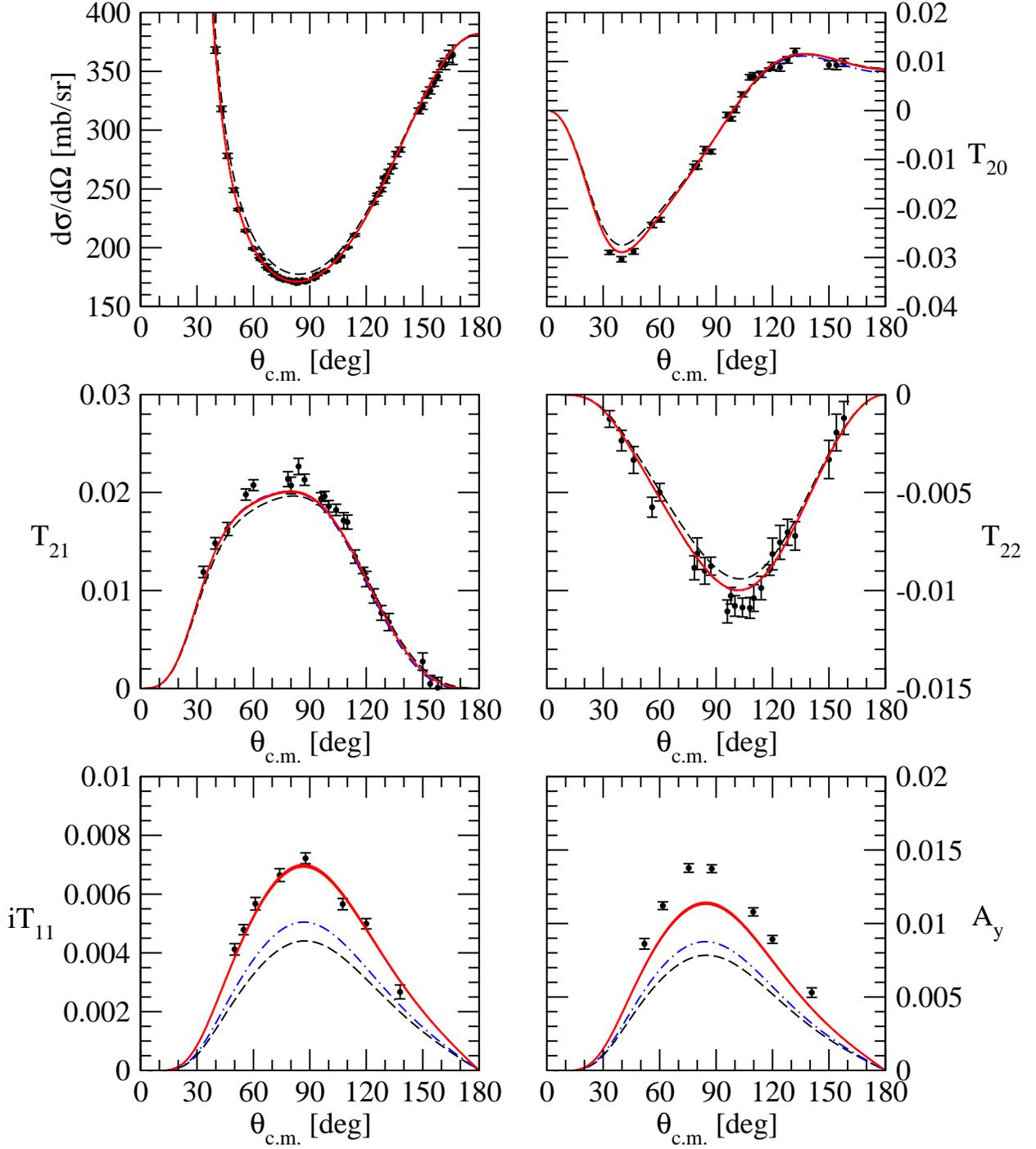}}
  \caption{(Color online)
    Predictions obtained with the three-nucleon interaction
    models discussed in the text with $\Lambda=200-500$ MeV
    (red bands) for a set of cross section and polarization
    $p-d$ observables at 0.666~MeV center-of-mass energy,
    as compared to the  purely two-body AV18 interaction
    (dashed black lines), to the 
    AV18/UIX two- and three-nucleon interaction (dashed-dotted blue lines),
    and to the experimental data of Ref.~\cite{Wood:2001}.
\label{fig:set06}}\end{figure}

With the interaction fitted using the $E_{c.m.}=2$ MeV data
of Ref.~\cite{Shimizu:1995}, we can perform a study at lower
energies, where experimental data exist. As a representative example we
show in Fig.~\ref{fig:set06} the results corresponding to $E_{c.m.}=0.666$~MeV,
from which we can observe that the adopted interaction captures quite nicely
the energy dependence of the data. In Ref.~\cite{Girlanda:2018},
a fit including all the subleading operators of Eq.~(\ref{eq:vcontact2})
leads to predictions in even better agreement with the data.
However, in order to obtain further improvements,
a global fit at multiple energies should be performed.

\subsection{$p-^3$He and $n-^3$H scattering}
\label{subsec:A4_scattering}
The study of $N-d$ scattering to constrain the three-nucleon force has the
limitation of being mostly restricted to the isospin $T=1/2$ channel. From this
perspective, $A=4$ systems open new possibilities, besides being of direct
relevance for the role they play in many reactions of astrophysical and
cosmological interest. The HH method has been used in this context to address
first of all the $n-^3$H~\cite{Viviani:2008} and
$p-^3$He~\cite{Viviani:2010,Viviani:2013} elastic scattering at
low energies. The HH method applied to these systems has been
benchmarked in Ref.~\cite{Viviani:2011} with the only two other
{\it ab-initio} methods which can study low-energy scattering states,
with full inclusion of the Coulomb interaction. They are
the AGS equations solved in momentum
space (see for a review Refs.~\cite{Fonseca:2017,Deltuva:2019} and references
therein),
and the FE method in configuration
space (see Ref.~\cite{Lazauskas:2019}. This topic is also covered
in the present Research Topic).
All these methods differ by less than 1\%, which is
smaller than the experimental
uncertainties of the available data. The agreement found using softer potentials
of the $V_{low-k}$-type is even better.

\begin{figure}[htb]
\centerline{\includegraphics[scale=0.5]{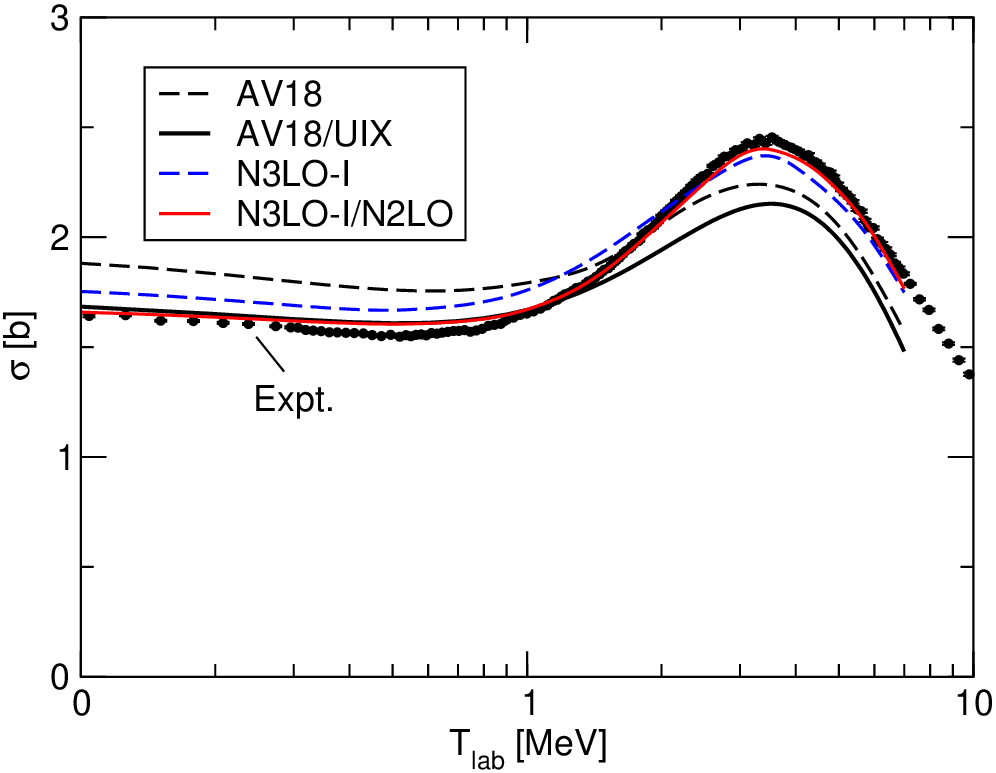}}
\caption{(Color online)
  $n-^3$H total cross sections calculated with the AV18 (dashed black
  line), AV18/UIX (solid black line), N3LO-I (dashed blue line),
  and the N3LO-I/N2LO (solid red line)
  potential
  models as
  function of the incident neutron laboratory energy $E_n$. The experimental
  data are from Ref.~\protect\cite{Phillips:1980}.}
\label{fig:tcs}
\end{figure}

The $n-^3$H elastic scattering total cross section
is shown in Fig.~\ref{fig:tcs}. From inspection of the figure,
we can see a sizable dependence
on the three-nucleon interaction,
both in the very low-energy region and in the peak region
(for neutron laboratory energy $E_n\sim 3.5$~MeV). Indeed,
at very low-energy, it is crucial to have a
correct description of the triton binding energy in order
to reproduce the data, whereas in the peak
region there is more model dependence.
The HH calculations of Fig.~\ref{fig:tcs}
have been performed using the non-local chiral N3LO-I
two-nucleon
potential, also supplemented by the chiral N2LO three-nucleon interaction
of Ref.~\cite{Navratil:2007} with
the LECs fixed to reproduce the $A=3,4$ binding energies. This
leads to a
remarkable agreement with the available experimental data in the low-energy
region. The chiral N3LO-I model seems to perform better than the AV18
also in the peak region.

\begin{figure}[htb]
\centerline{ \includegraphics[scale=0.5]{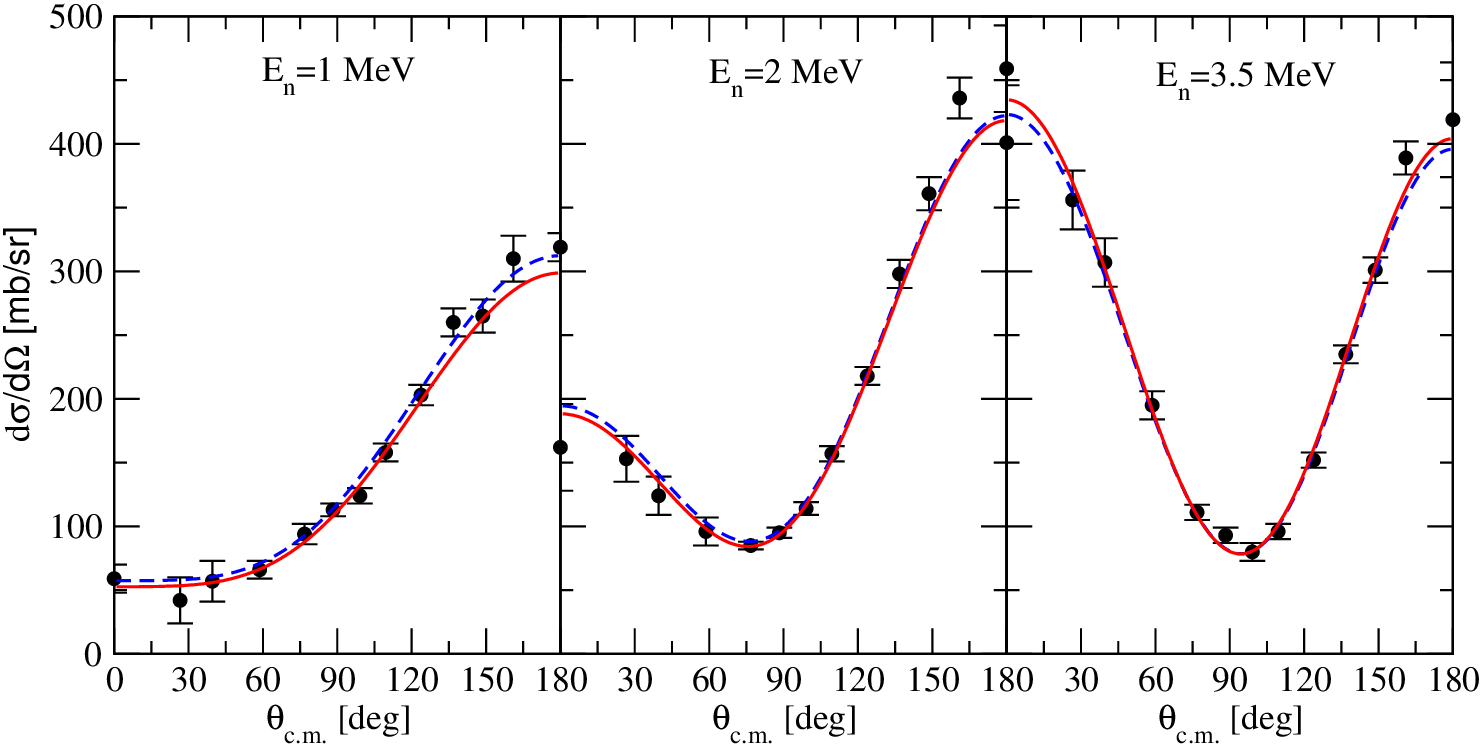}}
\caption{(Color online)
  $n-^3$H differential cross sections calculated with the
  N3LO-I (dashed blue lines) and the N3LO-I/N2LO (solid red lines) interaction
  models for three different incident neutron energies. The experimental data
  are from Ref.~\protect\cite{Seagrave:1960}.}
\label{fig:dcs}
\end{figure}

In Fig.~\ref{fig:dcs} we show the $n-^3$H
differential cross section compared 
to the experimental data at three different neutron
laboratory energies. As it is clear from inspection of the figure,
the N3LO-I/N2LO results are in nice agreement with the data.
A further study of convergence with respect to chiral orders
and of cutoff dependence would be highly desirable, and it is
currently underway.

\begin{figure}[hbt]
\centerline{  \includegraphics[scale=0.5]{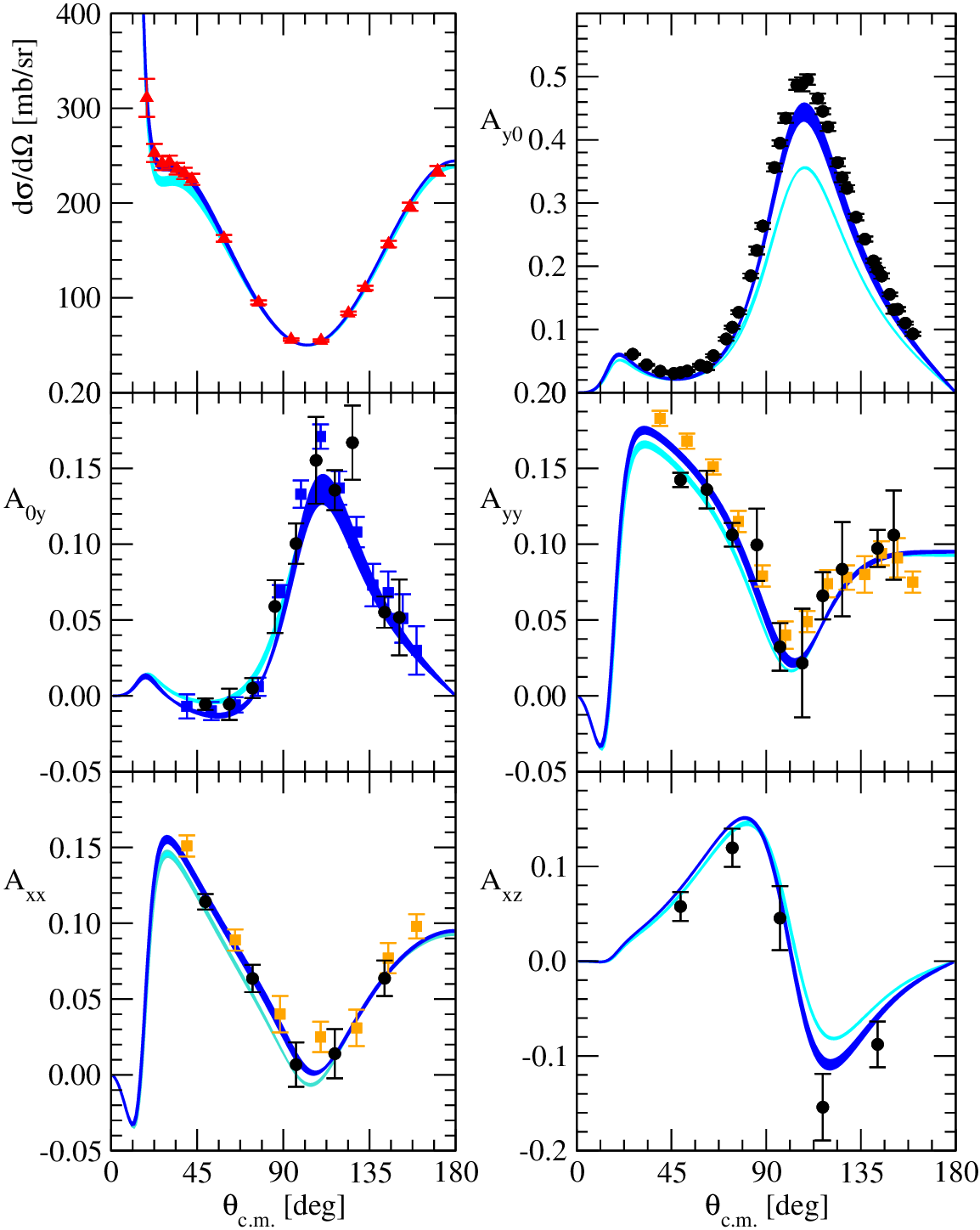}}
\caption{(Color online) $p-^3$He differential cross section, analyzing
  powers and various
  spin correlation coefficients at proton laboratory energy
  $E_p=5.54$ MeV, calculated with only the two-nucleon N3LO-I
  (light cyan band) or with two- and three-nucleon interaction
  N3LO-I/N2LO (darker
  blue band). The experimental data are from
  Refs.~\protect\cite{Fisher:2006,Alley:1993,Viviani:2001}.
  See text for more details.} 
\label{fig:6obs}
\end{figure}

Much more accurate data are available for $p-^3$He elastic scattering,
whose polarization observable have also been accurately
measured \cite{Daniels:2010}. Similarly to the $p-d$ case, there is a
strong discrepancy between theory and experiment for the proton analyzing
power $A_y$. In Ref.~\cite{Viviani:2013} the HH method has
been applied with the N3LO-I/N2LO chiral potential model,
in this case obtained 
with two different values of the momentum cutoff
$\Lambda=500, 600$~MeV~\cite{Machleidt:2011},
and two different procedures to fix the LECs
entering the three-nucleon interaction, i.e.\ either
reproducing the $A=3,4$ binding energies~\cite{Navratil:2007},
or reproducing the triton binding energy and Gamow-Teller matrix
element in tritium $\beta$-decay~\cite{Marcucci:2012}.
We show in Fig.~\ref{fig:6obs} the corresponding results for proton
laboratory energy of 
5.54~MeV, compared to experimental data. The two bands reflect the cutoff
dependence and the model dependence introduced by the LECs determinations.
As it is clear, the $A_y$ discrepancy is largely reduced down to the
8-10\% level. Note that these asymmetries are 10 times larger
in the $A=4$ systems than for $p-d$ and $n-d$.
The remaining discrepancy, although it appears small,
is of the order of 0.05, the size of $A_y$ for $p-d$. Therefore,
we expect that the subleading components of the three-nucleon
interactions discussed in Sec.~\ref{subsec:nd-scattering}
could give a correction of the necessary order of magnitude to solve
the remaining discrepancy. Work is in progress in this
direction.

\subsection{$p-^3$H and $n-^3$He scattering}
\label{subsec:A4-scattering2}
\begin{figure}[htb]
  \includegraphics[scale=0.7,clip,angle=0]{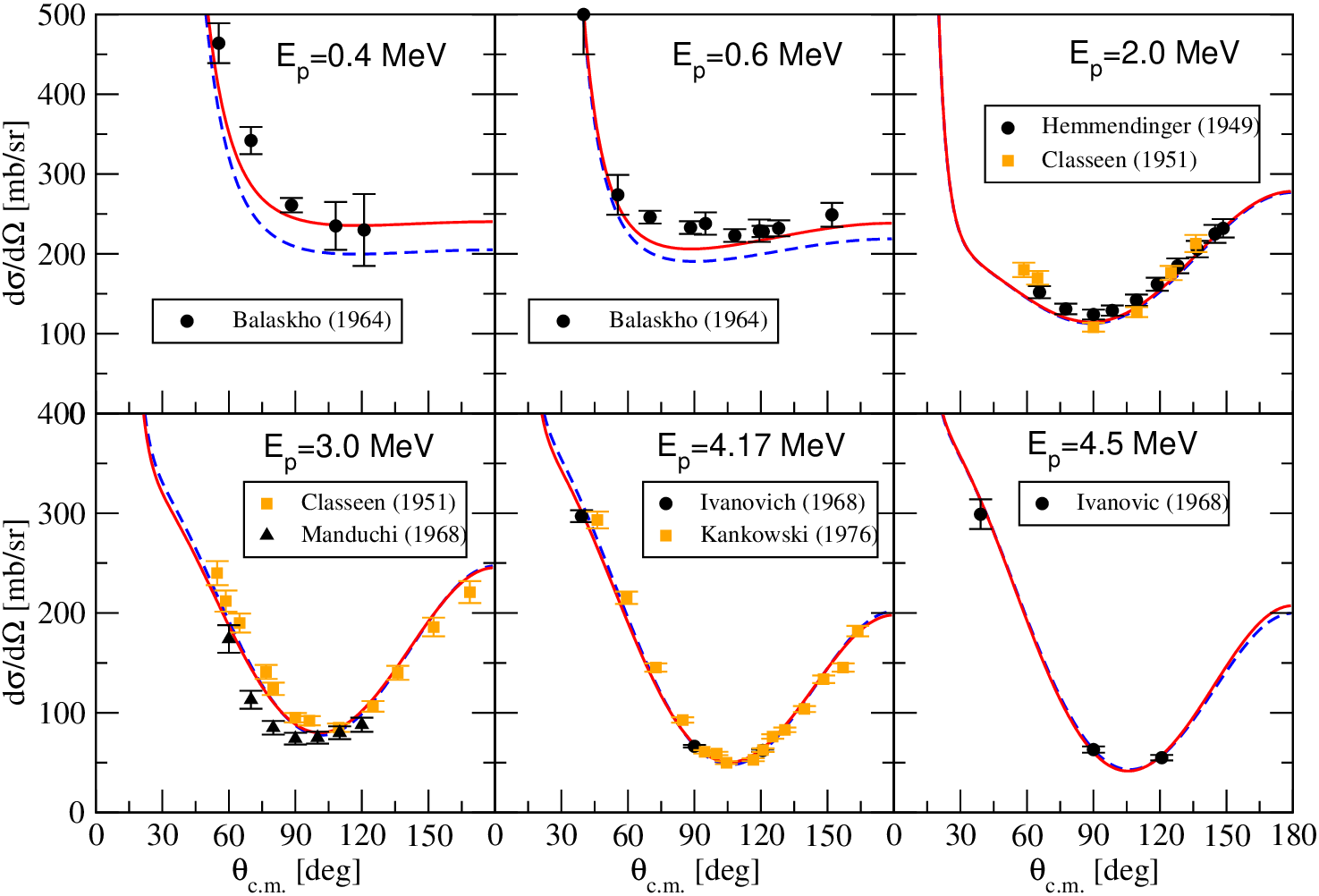}
  \caption{(Color online) $p-^3$H differential cross section at several values
    of the proton laboratory beam energy $E_p$,
    calculated with the N3LO-I (dashed blue lines) and with the N3LO-I/N2LO
    (solid red lines) interactions. The experimental data are from
    Refs.~\cite{Hemmendinger:1949,Claassen:1951,Balashko:1965,Manduchi:1968,
      Ivanovich:1968,Kankowsky:1976}.}
   \label{fig:ph3x}
\end{figure}

\begin{figure}[htb]
  \includegraphics[scale=0.7]{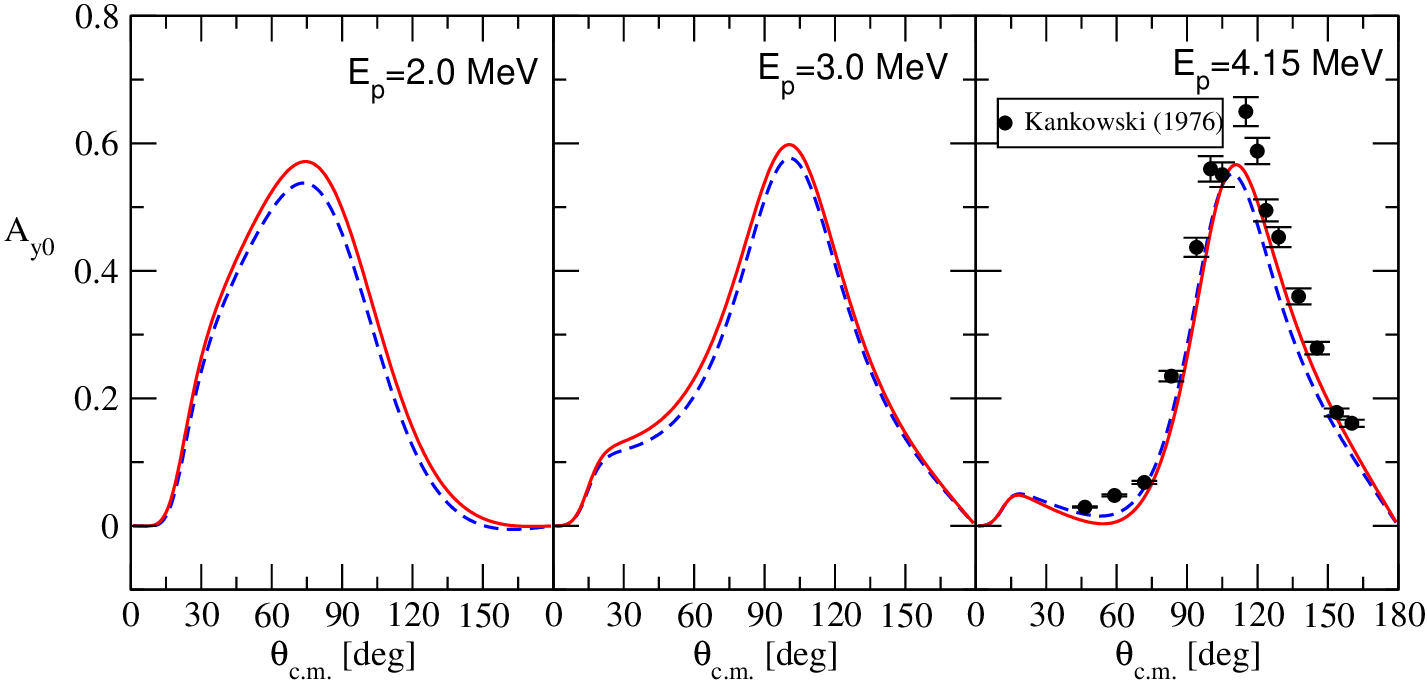}
  \caption{(Color online) $p-^3$H proton analyzing power at three values of
    the proton laboratory beam energy $E_p$ calculated with the N3LO-I
    (dashed blue lines) and with the N3LO-I/N2LO (solid red lines)
    interactions. The experimental data are from
    Ref.~\cite{Kankowsky:1976}.}
   \label{fig:ph3ay}
\end{figure}

The treatment of $p-^3$H and $n-^3$He scattering, even below
the $d+d$ threshold, is more challenging due
to the coupling between these two channels and to the presence of both
isospin 0 and 1 states. Also in this case, recently, in
Ref.~\cite{Viviani:2016}, a benchmark calculation
has been performed with the HH, AGS and FE methods, using the N3LO-I
interaction. Good agreement among the three
methods has been found, with discrepancies smaller than the uncertainties
in the experimental data. 
In Refs.~\cite{Viviani:2017,Viviani:2018}, we have studied with the HH
method the effect of the inclusion
of the N2LO three-nucleon interaction, with the LECs fixed from the triton
binding energy and the Gamow-Teller matrix element in the
tritium $\beta$-decay~\cite{Marcucci:2012}.
We show in Fig.~\ref{fig:ph3x} the $p-^3$H differential cross section,
for which, only at very low energies, below the opening of the $n-^3$He
channel, some sizable effects are
visible. Otherwise, the three-nucleon interaction contributions
are found very small.
The $p-^3$H analyzing power at three values of the laboratory beam energy
are shown in Fig.~\ref{fig:ph3ay}. Also for this observable,
the three-nucleon interaction effect is found too small
to improve the agreement with the available
experimental data.

\begin{figure}[hbt]
  \includegraphics[width=\columnwidth,clip,angle=0]{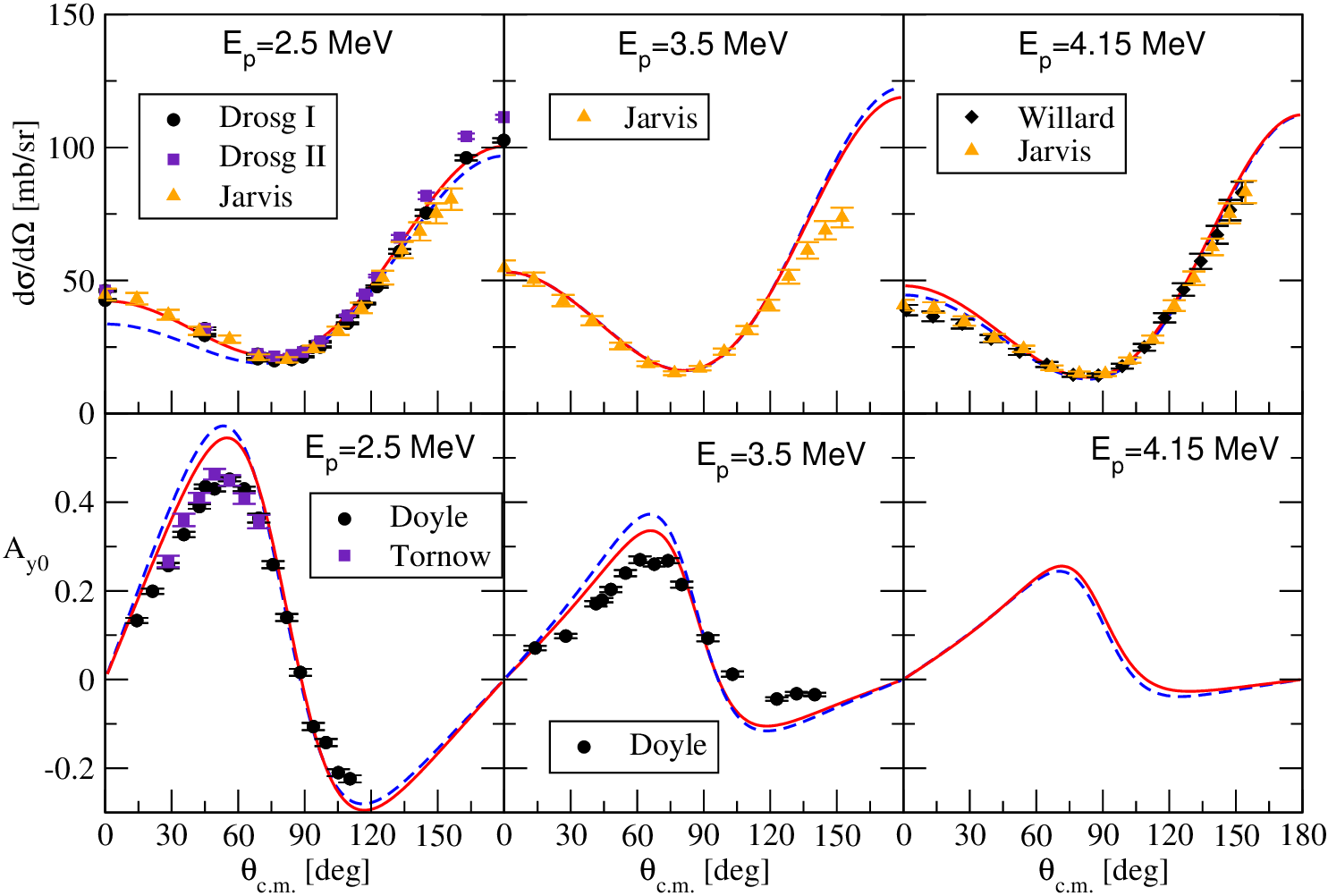}
  \caption{(Color online) $p+^3$H $\to n+^3$He differential cross section and
    proton analyzing power at three values of the proton laboratory beam
    energy $E_p$ calculated with the N3LO-I (dashed blue lines) and with
    the N3LO-I/N2LO (solid red lines) interactions. The experimental
    data are from
    Refs.~\cite{Willard:1953,Jarvis:1956,Drosg:1980,Doyle:1981,Tornow:1981}.}
   \label{fig:ph3-nhe3}
\end{figure}

We conclude showing in Fig.~\ref{fig:ph3-nhe3}
the HH results for the differential cross section and proton
analyzing power of the 
charge-exchange reaction $p+^3$H $\to n+^3$He at three different
proton laboratory energies, compared with the experimental data.
By inspection of the figure, we can see that also in this
case the effects of the three-nucleon interaction are quite small,
and sometimes go in the wrong direction as
compared to the experimental data, as for the analyzing power $A_{y0}$.
It is important to notice that this observable is mostly sensitive to
the two-nucleon interaction. Therefore it could be used
for a more stringent tests of the two-nucleon force.

\section{Conclusions and outlook}
\label{sec:concl-out}

In this work we have presented a review of the HH method,
focusing on the most significant achievements after the year 2008,
when the previous review on the HH method~\cite{Kievsky:2008}
was published. We have also included a presentation of the HH formalism
with some detail, in order to make the reader appreciate the main concepts
of the method and to provide him/her the instruments needed to
implement the method by him/herself. We have then focused on the 
latest results obtained within the HH method. We can summarize
the situation as follows: 
the HH method can solve the three- and four-body bound-state problem
with great accuracy and with essentially any (local and non-local)
model for the two-nucleon interaction available in the literature.
The three-nucleon interaction models used so far are
however only local. The $A=3,4$ scattering states have been studied with
any potential (again local and non-local)
below the target nucleus breakup threshold.
Using local potentials, also the elastic channel above
the breakup threshold have been investigated.
The HH method has then a wide range of applications: it has been used
not only to test the models for the two- and three-nucleon
interactions, but also to determine the parameters entering in the
subleading three-nucleon contact interaction, derived
in Ref.~\cite{Girlanda:2011}. This has allowed one to
construct a model for the three-nucleon interaction
able to solve, at least within the (preliminary) hybrid framework
of Ref.~\cite{Girlanda:2018},
some long-standing puzzles, as the $A_y$-puzzle.
Furthermore, the HH method has 
been widely used in the study of nuclear
reactions of astrophysical interest, as well as the electroweak
structure of light nuclei~\cite{Marcucci:2005,Marcucci:2016,Marcucci:2016b}.

The HH method has still a lot of potentialities, which will be explored
in the near future. First of all, we will implement the method
to the case of non-local three-nucleon interaction. This is widely
requested, in order to have consistency in the two- and three-nucleon
cutoff functions which appear in the models
derived within chiral effective field theory for instance
in Refs.~\cite{Entem:2003,Entem:2017}.
Once the LECs $c_D$ and $c_E$ will be determined
using the non-local three-nucleon interaction with the
same procedure outlined in Sec.~\ref{sec:res},
they will be used in {\it fully consistent} studies of other
systems, as nuclear and neutron matter.

Secondly, we can mention only preliminary applications of
the HH method to describe breakup reactions in $A=3$~\cite{Garrido:2014}.
Work on the implementation of the HH method to the breakup channels 
in $A=3,4$ is currently underway. It does not require
significant modifications of the method, but still it has not been
performed yet. Once done, the three- and four-body nuclear
systems will be completely covered by the method.

As mentioned above, the extension of the method to the $A=5,6$ nuclear
systems has been investigated and the first results
obtained using a $V_{low-k}$ interaction will appear
soon and are indeed very promising. This is a major step for the
HH method, as it will allow us to tackle a large number of
interesting subjects, and especially a large number of
nuclear reactions of astrophysical interest.
From a first investigation, the further extension of the method to even larger
values of $A$, i.e.\ $A=7,8$, seems feasible.

Finally, in order to have access to higher mass nuclei,
we could take advantage of the strong clusterization present
in some of them, as, for instance, in $^9$Be, which can be
studied as a $\alpha-\alpha-n$ system. In order to do so, 
the HH method must then be extended to the case of non-equal
mass systems. And this, in turn, will allow to study also
more exotic systems, as hypernuclei, where one
nucleon is replaced with an hyperon. Works
along this line have started in Ref.~\cite{Nannini:2018},
and are conducted also by other groups~\cite{FerrariRuffino:2017}.

In conclusion, the HH method has quite a ``glorious'' history, and has
fulfilled its service in the continous test of the nuclear interaction
models.
However, this service is not yet at an end, and we expect to
see the HH method playing a protagonist role also in the next years.

\section*{Funding}
The work of J.D.E. was supported by the Fonds de la Recherche
Scientifique (FNRS) under Grant Number 4.45.10.08.

\section*{Acknowledgments}
The computational resources of the Istituto Nazionale di Fisica
Nucleare (INFN), Sezione di Pisa, are gratefully acknowledged.

\bibliography{HHmethod.arX}


\end{document}